\documentclass[12pt]{article}

\usepackage{epsfig}
\usepackage{amsmath,pifont}
\usepackage{amssymb,amsxtra}
\usepackage{moreverb}
\usepackage{xspace}
\usepackage{subfigure}
\usepackage{hyperref}
\usepackage{comment}
\usepackage{natbib}
\usepackage{color, soul}

\newcommand{\PetFMM}{\textsc{p}et\textsc{fmm}\xspace}
\newcommand{\PetRBF}{\textsc{p}et\textsc{rbf}\xspace}
\newcommand{\parmetis}{\textsc{p}ar\textsc{metis}\xspace}

\newcommand{\Kern}{\mathbb{K}}
\newcommand{\ME}{\textsc{me}\xspace}
\newcommand{\LE}{\textsc{le}\xspace}
\newcommand{\SUMAN}{\textsc{sum-an}\xspace}
\newcommand{\SUMGone}{\textsc{sum-g1}\xspace}
\newcommand{\SUMGtwo}{\textsc{sum-g2}\xspace}
\newcommand{\FEMD}{\textsc{fem-d}\xspace}
\newcommand{\FEMGT}{\textsc{fem-gt}\xspace}
\newcommand{\mtx}[1]{\mathbf{#1}}
\newcommand{\vtr}[1]{\mathbf{#1}}

\title{Optimal, scalable forward models for computing gravity anomalies}
\author {D.A. May$^1$, M.G. Knepley$^2$\\ 
\small $^1$ Department of Earth Sciences, ETH Z\"urich, Z\"urich, Switzerland. \\
\small E-mail: dave.mayhem23@gmail.com \\
\small $^2$ Computation Institute, University of Chicago, Illinois, USA. \\
\small E-mail: knepley@ci.uchicago.edu
}
\date{}

\begin{document}
\maketitle

\begin{abstract}
	We describe three approaches for computing a gravity signal from a density anomaly. The first approach consists of the classical ``summation'' technique, whilst the remaining two methods solve the Poisson problem for the gravitational potential using either a Finite Element (FE) discretization employing a multilevel preconditioner, or a Green's function evaluated with the Fast Multipole Method (FMM). The methods utilizing the Poisson formulation described here differ from previously published approaches used in gravity modeling in that they are optimal, implying that both the memory and computational time required scale linearly with respect to the number of unknowns in the potential field. Additionally, all of the implementations presented here are developed such that the computations can be performed in a massively parallel, distributed memory computing environment. Through numerical experiments, we compare the methods on the basis of their discretization error, CPU time and parallel scalability. We demonstrate the parallel scalability of all these techniques by running forward models with up to $10^8$ voxels on 1000's of cores.
\end{abstract}
% Keywords: gravity anomaly -- forward modeling -- summation -- multigrid -- fast multipole method.

% ====================================================== %
\section{Introduction} \label{SEC_Introduction}
\subsection{Background} \label{SSEC_Background}
\begin{figure}
\begin{center}
	\includegraphics[width =90mm]{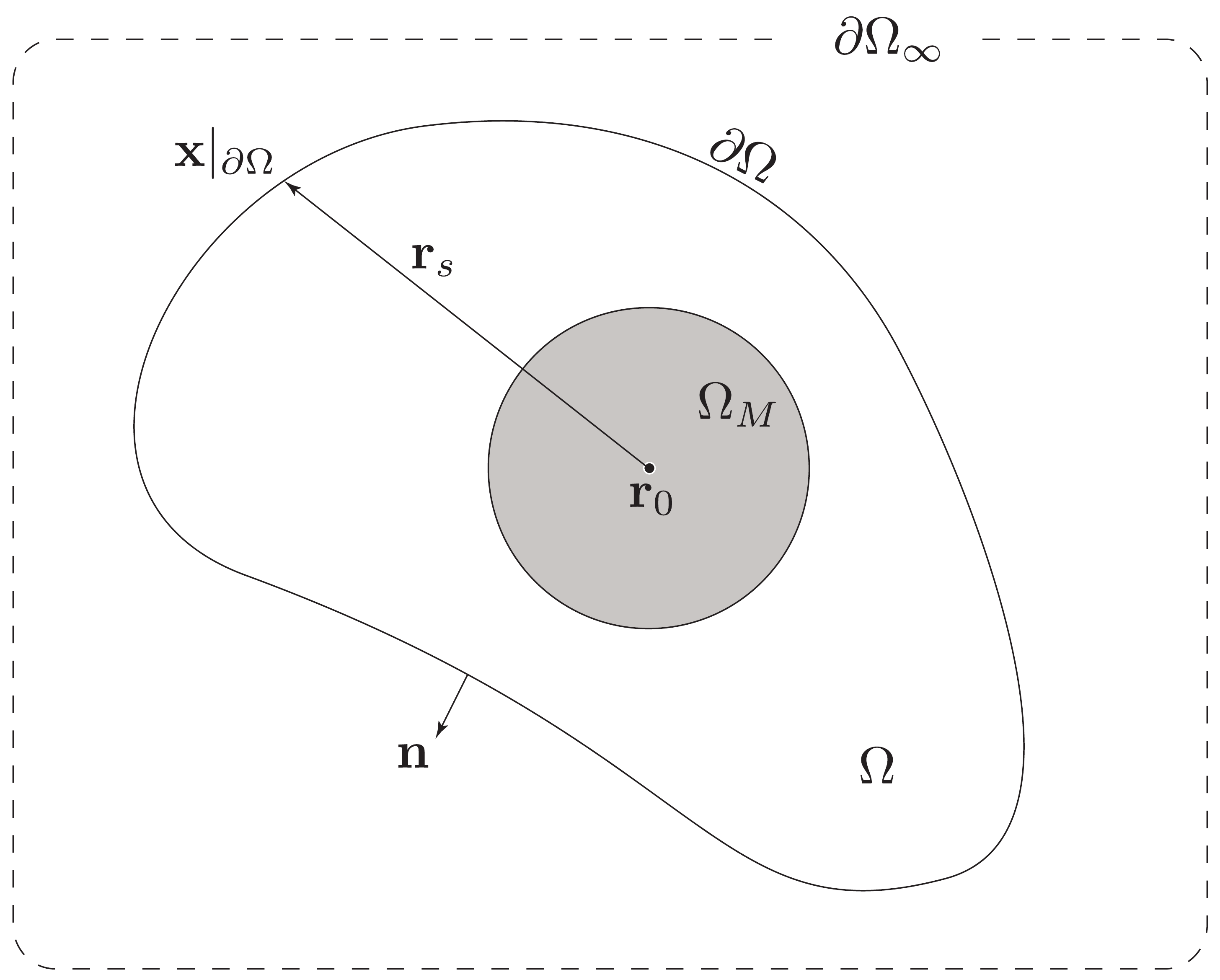}
	\caption{Problem domain for computing gravity. Here we denote the infinite domain boundary by $\partial \Omega_{\infty}$, the model domain by $\Omega$ and density anomaly domain by $\Omega_M$. The center of mass of $\Omega_M$ is denoted by $\vtr r_0$ and $\vtr n$ is the outward pointing normal to the boundary of $\Omega$.}
	\label{FIG_domain}
\end{center}
\end{figure}

The use of forward models to compute synthetic gravity signals is necessary to conduct inversions of the subsurface density structure.
Given a volume $\Omega_M$ over which we have a density field $\rho(\vtr x)$, the gravity attraction at a point $\vtr r = (r,s,t)$ due to this body can be computed via 
\begin{equation}\label{eq:gravity}
	\vtr g(\vtr r) = G \int_{\Omega_M} \rho( \vtr x) \frac{\vtr r -\vtr x}{ [ (r-x)^2 + (s-y)^2 + (t-z)^2 ]^{3/2}  } \, dV.
%	g_z(\vtr r) = G \int_{\Omega_M} \rho( \vtr x) \frac{t -z}{ [ (r-x)^2 + (s-y)^2 + (t-z)^2 ]^{3/2}  } \, dV.
\end{equation}
An alternative way to compute the gravity field is to solve the gravitational potential equation
\begin{equation}
	\nabla^2 \phi = -4 \pi G \rho( \vtr x ) \qquad \text{in} \medspace \medspace \Omega_{\infty},
	\label{EQ_gravpotential}
\end{equation}
where $\phi$ is the potential, $G$ is the gravitational constant, $\Omega_{\infty}$ denotes the entire free space and we assume that $\rho(\vtr x)=0, \medspace \forall \medspace \vtr x \notin \Omega_M$. The potential is subject to the following Dirichlet boundary condition
\begin{equation}
	\phi = 0, \qquad \text{at} \medspace\medspace \vtr x = \infty .
	\label{EQ_gravzeroinfinity}
\end{equation}
The gravity field sought is given by the gradient of the potential $\phi$;
\begin{equation}
	\vtr g(\vtr x) = - \nabla \phi.
	\label{EQ_potentialdefinition}
\end{equation}
The physical model is depicted in Fig.~\ref{FIG_domain}.
Forward gravity models typically fall into one of two categories: summation based techniques which evaluate Eq.~\eqref{eq:gravity}, or partial differential equation (PDE) based techniques which solve the gravitational potential formulation in Eqns.~\eqref{EQ_gravpotential}-\eqref{EQ_potentialdefinition}. 
%%%

The summation methods require the subsurface density structure to be discretized into a set of volumes. At each location $\vtr r$, in the model domain where a gravity signal is sought, the gravitational contribution from each density element in the domain is evaluated using Eq.~\eqref{eq:gravity} and summed. The summation methods differ in the manner in which the integral expression in Eq.~\eqref{eq:gravity} is evaluated. Several analytic approaches exist in which a closed form expression for Eq.~\eqref{eq:gravity} is used in either Cartesian (see \cite{li.chou.339.1998} for an overview) or spherical coordinates \citep{john.lith.6999.1972,smit.etal.783.2001}. The limitation of analytic expression is that one is forced to choose a spatial discretization for the density structure which is orthogonal to the coordinate system, and the density is usually required to be constant over each element. The complexities and discretization restrictions of the analytic method can be overcome by using a sufficiently accurate quadrature scheme to approximate Eq.~\eqref{eq:gravity}. This approach permits any spatial discretization to be used provided a high accuracy quadrature rule can be defined over the geometry of each cell used in the discretization~\citep{asgh.fres.1.2007}.

Recently there has been some interest in using PDE based approaches to compute gravity anomalies, as these methods have been demonstrated to be both faster and produce more accurate forward models than the summation techniques. In \cite{yong.wang.696.2005}, a finite element method was used to obtain the solution to the Poisson equation. They favoured the finite element method over the finite difference method as the former allowed more geometric freedom in meshing the density anomalies and the formulation easily permitted a variable density field within each voxel. Their formulation utilized a Robin type boundary condition to approximate the boundary condition in Eq.~\eqref{EQ_gravzeroinfinity}. The method was regarded as being ``fast'' since within a finite size domain, the Robin condition yielded a smaller error than setting $\phi=0$ on the boundary of a finite domain. That is, the convergence of the error using this method was faster than simply setting $\phi = 0$ on the boundary of the finite domain.
%In contrast, ~\cite{farq.mosh.417.2009} employed a finite difference discretization to solve Eq.~\eqref{EQ_gravpotential}. In this work, the Poisson boundary condition is approximated by ensuring that the model boundaries are far away from the density anomaly to mimic the condition, $\phi = 0$ at $\vtr x=\infty$. 
In contrast, ~\cite{farq.mosh.417.2009} employed a finite difference discretization to solve Eq.~\eqref{EQ_gravpotential}, where the boundary condition $\phi = 0$ at $\vtr x=\infty$ is approximated by ensuring that the model boundaries are ``far'' from the density anomaly, which in their work constituted using a model domain with side lengths six times larger than the side length of the anomaly.

%% forward models 
%Within the summation approaches, every cube in the discretisation makes a gravity contribution to every station. If we expressed the gravity integral in matrix form, as $g_i = A_{ij} \rho_j$, the matrix $A$ is dense. For high resolution 3D inversions, the storage required for $A$ and its associated covariance matrix quickly becomes unmanageable.

The development of fast and efficient forward models is crucial to enable high resolution inversion to be performed. In considering the computation complexity of the summation algorithm, we see that if we discretize the domain with $N$ density elements and we have $M$ measurements, i.e. locations where we will evaluate the gravity, the calculation will require $\mathcal O(M N)$ time. 
%%
%Given the ease with which gravity measurements can now be made, either directly from the field using an absolute or %indirectly via tomographic techniques %\hl{(I think this statement is bullshit)}, 
%applied geophysics studies may typically have values of $M$ on the order of 10,000. 
%%
Given the ease with which gravity measurements can be made on a regional scale using either a land-based relative gravimeter or via airborne measurements, or on a global scale using satellite based gravimetry, applied geophysics studies may typically have values of $M$ on the order of 10,000. 
The number of measurements $M$ is continually increasing as new techniques are developed, or existing techniques become affordable or automated. We note that the computational cost of evaluating the gravity contribution from one element via Eq.~\eqref{eq:gravity} is not insignificant. Even the simplest 1-point quadrature rule requires: 5 additions, 7 multiplications and one square root, which is equivalent to the cost of $\sim20$ multiplications \citep{agner_webpage}.
%The resulting expressions obtained from the analytic approach are also typically expensive to evaluate numerically. For example, in cartesian coordinates the analytic expression of $g$ involves operations such as arctan and the natural logarithm, which may take as much time to evaluate as many tens of multiplications.

Using the PDE approach, one obtains the value of the potential over the entire domain, from which the gravity can be computed as a post-processing task. Consequently, the PDE approaches have a computational complexity which is not a strong function of the number of evaluation points, but instead is dominated by the complexity of the linear solver ($X$) used to obtain the potential, i.e. the overall method scales according to $\mathcal O(N + X)$. If sparse direct factorizations (such as Cholesky or LU) are used, the solve time will scale like $X = \mathcal O(n^{3/2})$ in 2D and $X = \mathcal O(n^2)$ in 3D, where $n$ is the number of unknowns used to represent the discrete potential field. The memory usage for these solvers is $\sim \mathcal O( n \log n)$ and $\sim \mathcal O(n^{4/3})$ for 2D and 3D respectively \citep{Li.Widl.1415.2007}. If unpreconditioned Krylov methods like conjugate gradient are used, the solve time will scale according to $X = \mathcal O(n^{3/2})$ and $\mathcal O(n^{4/3})$ in 2D and 3D respectively. Numerous optimal multilevel preconditioners exist for the Poisson equation in which both the solve time and memory usage will scale like $\mathcal O(n)$ \citep{trot.etal.2001}.

%[(regional) Terresterial gravimetry techniques, airborne gravimetry and (global) satellite based gravimetry]

\subsection{Present work}
Here, we examine several variants of the summation method, a finite element method with two types of boundary conditions and a fast multipole method to compute synthetic gravity fields. Our examination of the different methods focuses on the accuracy and the algorithmic complexity (optimality) of the techniques. All of the methods used in this study are developed to be executed on massively parallel, distributed memory computer architectures. We also examine the parallel performance (scalability) of the three classes of the methods under consideration.

% ====================================================== %
\section{Numerical techniques} \label{SEC_Techniques}

	\subsection{Summation} \label{SSEC_Summation}
We considered three variants of the summation technique in this study. Each of the summation techniques is defined in a Cartesian coordinate system and utilized a structured mesh of hexahedral cells to discretize the density field. The model domain considered was always ``brick'' shaped and thus was easily decomposed into a set of $M_x \times M_y \times M_z$ cells .
Within each cell, the density is assumed to be constant. 
%The gravity from Eq.~\eqref{eq:gravity} is evaluated at the centroid of each hexahedral. 
The first summation approach (which we identify as \SUMAN) uses the analytic expression from \cite{li.chou.339.1998} to  evaluate the vertical component of the gravitational contribution $g_z(\vtr x)$, given by Eq.~\eqref{eq:gravity}. The other two methods we consider use either a one point Gauss (\SUMGone),  or a two point Gauss (\SUMGtwo) quadrature scheme to evaluate the gravity integral. 

Parallelism is achieved in the summation methods via a spatial decomposition of the mesh used to discretize the density field. The locations where the gravity field is required to be evaluated are duplicated on each processor. Every processor calculates a local gravitational contribution at each evaluation point from a subset of cells within the entire domain. This operation can be completed without any communication. The only communication required is a global reduction of the local gravity contributions from each processors local subdomain.

	\subsection{Finite element method} \label{SSEC_FEM}
The Poisson equation in Eq.~\eqref{EQ_gravpotential} is solved using a standard Galerkin Finite Element (FE) formulation \citep{hugh.208.1987}.
The variational form is given by
\begin{equation}
	\int_{\Omega_\infty} \!\! v \nabla^2 \phi \, dV = 4 \pi G \int_{\Omega_\infty} \!\! v \rho(\vtr x) \, dV,
	\label{EQ_varform}
\end{equation}
where $v$ is a test function which vanishes on all Dirichlet boundaries.
Applying integration by parts to the second order derivative in Eq.~\eqref{EQ_varform}, we obtain
\begin{equation}
	-\int_{\Omega_\infty} \!\! \nabla v \centerdot \nabla \phi \, dV
	+ \int_{\partial \Omega_\infty} \!\! v \nabla \phi \centerdot \vtr n \, dS
	= 4 \pi G \int_{\Omega_\infty} \!\! v \rho(\vtr x) \, dV.
	\label{EQ_varform2}
\end{equation}

Here we consider using two different approaches to approximate the ``Dirichlet at infinity'' boundary condition in Eq.~\eqref{EQ_gravzeroinfinity}. Both methods first approximate the entire free space domain $\Omega_\infty$, by a finite sized domain $\Omega$, satisfying $\Omega_M  \subseteq \Omega$. %$\Omega >> \Omega_M$.
The first approximation of Eq.~\eqref{EQ_gravzeroinfinity} we consider simply requires that
\begin{equation}
	\phi \, \rvert_{\partial \Omega} = 0,
\end{equation}
where $\partial \Omega$ denotes the boundary of $\Omega$.
Clearly, the larger the domain $\Omega$ is compared to the domain of the density anomaly $\Omega_M$, the better the approximation. We will denote this particular boundary condition approximation as \FEMD.

The second approximate boundary condition we considered was introduced by  \cite{yong.wang.696.2005} and consists of approximating the far field gravitational attraction on a finite sized domain $\Omega$. The far field gravity is approximated according to
\begin{equation}
	\vtr g \, \rvert_{\partial \Omega} = \frac{\phi}{\vtr r_s} \, \bigg\rvert_{\partial \Omega},
	\label{EQ_gravapprox}
\end{equation}
where $\vtr r_s = \vtr x \rvert_{\partial \Omega} - \vtr r_0$ and $\vtr r_0$ is the centroid of the density anomaly domain $\Omega_M$. These quantities are indicated on Fig.~\ref{FIG_domain}.
Using the definition of the potential from Eq.~\eqref{EQ_potentialdefinition}, we can introduce Eq.~\eqref{EQ_gravapprox} naturally into the variational problem in Eq.~\eqref{EQ_varform2} as a Robin boundary condition. We denote this boundary condition approximation as \FEMGT. For a thorough description of the finite element formulation and the implementation of the Robin boundary conditions, we refer readers to \cite{yong.wang.696.2005}.

As in the summation method, the domain consisted of a brick like geometry and was discretized with $M_x \times M_y \times M_z$ hexahedral elements. The discrete solution for $\phi$ was represented with piecewise trilinear ($Q_1$ basis) functions  over each hexahedral element. The same mesh was used to define the density structure. In the FE implementation used here, the density was assumed to be constant over each element. The resulting discrete problem from the FE discretization yields the sparse matrix problem
\begin{equation}
	\left[ \mtx L + \mtx F \right] \vtr x = \vtr b,
	\label{EQ_discretepoisson}
\end{equation}
where $\vtr x, \vtr b$ represent the discrete potential and force term, $\mtx L$ is the discrete Laplacian and $\mtx F$ is the term associated with the far field boundary condition appearing in the surface integral in Eq.~\eqref{EQ_varform2}. We note that $\mtx F =\mtx 0$ when the \FEMD approach is used. 

Following the solution of Eq.~\eqref{EQ_discretepoisson}, we compute the gravity within each element by interpolating the gradient of the trilinear basis functions used to approximate $\phi$. This approach has the disadvantage that the gravity field computed is discontinuous across element boundaries. The reconstruction of a continuous $C^0$ nodal field from the gradient of a finite element solution is a thoroughly studied problem. The Super Convergent Patch Recovery (SPR) \citep{zien.zhu.207.1992} and the Recovery by Equilibrium of Patches (REP) \citep{boro.zien.137.1997} are both appropriate techniques to recover an accurate nodal gravity field. In \cite{yong.wang.696.2005}, a nodal gravity field was computed using a global $L_2$ projection. A local $L_2$ projection can also be used \citep{hugh.208.1987}, which has the advantage of not requiring the solution of a global matrix problem. In practice, to enable the gravity field to be evaluated everywhere, a continuous gravity field defined on the nodes of the finite element mesh is the most useful representation. In this work however, we only use the results of the gravity field to compute error norms, for which the element wise, discontinuous representation of the gravity field is sufficient.

The matrix problem in Eq.~\eqref{EQ_discretepoisson} was solved using FGMRES \citep{saad.2003}, preconditioned with one V-cycle of geometric multrgrid (GMG). The GMG preconditioner we used is fairly standard and we refer to \cite{brig.etal.2000, wess.1992} and \cite{trot.etal.2001} for an introduction to these methods. Here we briefly summarize the components used in our multgrid preconditioner. 

The multigrid method utilizes a mesh hierarchy consisting of $n_l$ levels. Each level in the hierarchy defines a mesh of different spatial resolution. In the results presented here, a grid refinement factor of two was used between each grid level. The mesh at level $n_l$ has the finest resolution and represents the mesh used to discretize the potential field problem. The operator $\mtx A = \mtx L + \mtx F$ was defined on each mesh within the hierarchy by re-discretizing the PDE. Trilinear interpolation was used to define the restriction operator $\mtx R$, which is required to project nodal fields from a fine grid, to the next coarsest grid. Interpolation of fields from a coarse to fine grid was given by $\mtx R^T$.  On every grid level except the coarsest, we employed $N_k$ Richardson's iterations, combined with a Jacobi preconditioner as our smoother.  Given a vector $\vtr y_{k}$ at iteration $k$, the application of the smoother is given by the following sequence 
\begin{equation}
	\vtr y_{k+1} = \vtr y_k + \text{diag}( \mtx A )^{-1} \left( \vtr b - \mtx A \vtr y_{k} \right).
	\label{EQ_smoother}
\end{equation}
Unless otherwise stated, $N_k = 2$ was used in all experiments. On the coarsest grid level, the smoother was defined via an LU factorization. 

In our Poisson solver, the action of $\vtr A \vtr y_{k}$, required by the smoother in Eq.~\eqref{EQ_smoother} (on all grid levels expect the coarsest) and during each FGMRES iteration (finest grid only), was defined in a matrix-free manner. %using element-by-element matrix vector products. 
Similarly, $\text{diag}( \mtx A )$ was computed element-by-element, without explicitly assembling the full stiffness matrix $\mtx A$. On the coarsest grid, $\mtx A$ was explicitly assembled to allow an LU factorization to be performed.

At each iteration $i$ of the Krylov method, we monitor the 2-norm of the residual $\vtr r_i = \vtr b - \mtx A \vtr x_i$. The current estimated solution $\vtr x_i$ obtained from the iterative method was deemed to be converged if $\| \vtr r_i \|_2 < 10^{-10} \| \vtr r_0 \|_2$, where $\vtr r_0$ is the initial residual.  

Support for parallel linear algebra, Krylov methods and the structured mesh representation were provided by the Portable Extensible Toolkit for Scientific (c)omputation (PETSc) \citep{petsc-user-ref}. 

%In the parallel computations considered, the LU factorisation used on the coarse grid was performed using the parallel, multi-frontal method provided by the MUMPS package \citep{ame.etal.41.2001}. \hl{I use TFS in some runs...}

%Solver consists of a Krylov method preconditioned with a multilevel method (ether algebraic MG of geometric MG).

	\subsection{Fast multipole method} \label{SSEC_FMM}
The Fast Multipole Method (FMM) is an algorithm that accelerates the solution of an $N$-body problem,
\begin{equation}\label{eq:interaction}
  \vtr g(\vtr{x}'_j) = \sum_{i=1}^N \rho_i \, \Kern(\vtr{x}'_j, \vtr{x}_i),
\end{equation}
which is simply a discrete form of Eq.~\eqref{eq:gravity}. Here, $\vtr g(\vtr{x}'_j)$ represents the gravitational field evaluated at a point $\vtr{x}'_j$, where the field is generated by the influence of sources located at the set of points $\{\vtr{x}_i\}$. The sources are often associated with particle-type objects, such as charged particles, or in this case rock masses. In summary: $\{\vtr{x}'_j\}$ is a set of evaluation points;  $\{\vtr{x}_i\}$ is a set of source points with densities given by $\rho_i$; and $\Kern(\vtr{x}',\vtr{x})$ is the kernel that governs the interactions between evaluation and source particles. The kernel for the gravitational interaction in three dimensions is given by
\begin{equation}
  \Kern(\vtr{x}'_j, \vtr{x}_i) = \frac{\vtr{x}' - \vtr{x}}{\left| \vtr{x}' - \vtr{x} \right|^3}.
\end{equation}
Obtaining the field $\vtr g$ at all the evaluation points requires in principle $\mathcal{O}(M N)$ operations, for $N$ source points and $M$ evaluation points. The fast multipole method obtains $\vtr g$ approximately with a reduced operation count, $\mathcal{O}(M + N)$. 

\begin{figure}
\centering
\subfigure[Domain decomposition.]{\label{fig:spatialDecomposition-a}\includegraphics[width=140mm]{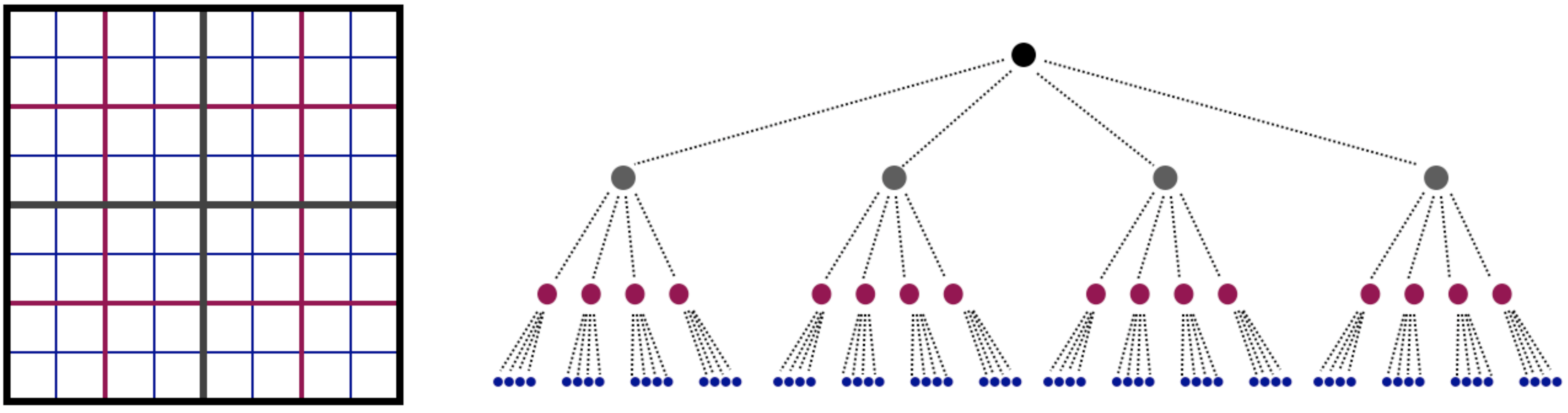}}
\subfigure[Near and Far field.]{\label{fig:spatialDecomposition-b}\includegraphics[width=140mm]{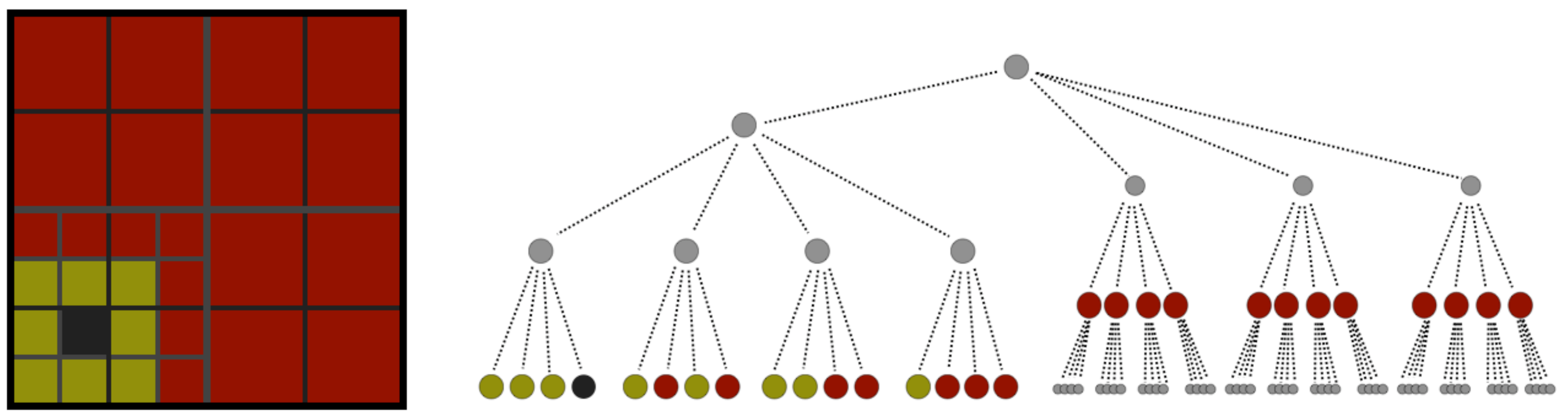}}
\caption{Quadtree decomposition of a two-dimensional domain: (a) presents a hierarchical tree related to the full spatial decomposition of the domain; (b) presents a colored two-dimensional spatial decomposition for interacting with particles in the black box, and its equivalence on the tree. The near-field is composed by the dark yellow boxes and the black box itself, while the far-field is composed by the dark red colored boxes. Notice that the far-field is composed of boxes of different levels of the tree structure. The relationships between the nodes of the tree simplify the process of composing the near and far domains.}
\label{fig:spatialDecomposition}
\end{figure}

In the FMM algorithm, the influence of a cluster of particles is approximately represented by a series expansion, which is then used to evaluate far-away interactions with controllable accuracy. To accomplish this, the  computational domain is hierarchically decomposed, allowing pairs of subdomains to be grouped into \textit{near} and \textit{far}, with far interactions treated approximately. Fig.~\ref{fig:spatialDecomposition} illustrates such a hierarchical space decomposition for a two-dimensional domain, associated to a quadtree structure. 

Using this decomposition of the computational domain, the sum in Eq.~\eqref{eq:interaction} can be decomposed as
\begin{equation}
  \vtr g(\vtr{x}'_j) = \sum_{k = 1}^{N_{\mathrm{near}}} \rho_{k} \Kern(\vtr{x}'_j , \vtr{x}_k) + \sum_{k = 1}^{N_{\mathrm{far}}} \rho_{k} \Kern(\vtr{x}'_j , \vtr{x}_k).
\label{eq:nearFarDecomposition}
\end{equation}
The first term, corresponding to the near field of an evaluation point, will have a small fixed size independent of
$N$. The second sum of Eq.~\eqref{eq:nearFarDecomposition}, representing the far field, will be evaluated efficiently using a series approximation so that the total complexity for the evaluation is $\mathcal{O}(N)$. We will use the following terminology for our field approximations:
\begin{description}
\item[Multipole Expansion] (\ME):  is a $p$ term series expansion that represents the influence of a cluster of particles at distances large with respect to the cluster radius.
\item[Local Expansion] (\LE): is a $p$ term series expansion, valid only inside a subdomain, used to efficiently evaluate a group of \ME s locally in a cluster of evaluation points.
\end{description}
The center of the series for an \ME is the center of the cluster of source particles, and it converges only outside a
given radius centered at the cluster of particles. In the case of an \LE, the series is centered near an evaluation
point and converges only inside a given radius.

The introduction of a single representation for a cluster of particles, via the multipole expansion, effectively permits a decoupling of the influence of the source particles from the evaluation points.  This is a key idea, resulting in the factorization of the computations of \ME's that are centered at the same point, so that the kernel can be written
\begin{equation}\label{eq:decouple}
  \Kern(\vtr{x}'_j, \vtr{x}_i) = \sum_{m = 0}^{p} C_{m}(\vtr{x}_i) f_{m}(\vtr{x}'_j)
\end{equation}
This factorization allows pre-computation of terms that can be reused many times, reducing the complexity of evaluation from $\mathcal{O}(N^2)$ to $\mathcal{O}(N \log N)$. Similarly, the local expansion is used to decouple the influence of an \ME from the evaluation points.  A group of \ME's can be factorized into a single \LE, which allows the $\mathcal{O}(N \log N)$ complexity to be further reduced to $\mathcal{O}(N)$. By representing \ME's as \LE's one can efficiently evaluate the effect of a group of clusters on a group of evaluation points.

\noindent{\bf Hierarchical space decomposition}

In order to make use of the \ME and \LE, the domain must be decomposed into near and far subdomain pairs. A hierarchical decomposition provides an efficient implementation for this operation. The hierarchical subdivision of space is associated to a tree structure (\emph{quadtree} structure in two dimensions, or an \emph{octree} structure in three dimensions) to represent each subdivision. 
%In fact, the \PetFMM provides a tree structure which works in any dimension. % Why is petfmm mentioned here? - it hasn't been introduced yet
The nodes of the tree structure are used to define the spatial decomposition, and different scales are obtained by looking at different levels. Consider Fig.~\ref{fig:spatialDecomposition-a}, where a quadtree decomposition of the space is illustrated. The nodes of the tree at each level cover the entire domain. The domain covered by a parent box is further decomposed into smaller subdomains by its child nodes. As an example of its use in FMM, consider Fig.~\ref{fig:spatialDecomposition-b} where the near-field for the \emph{black} colored box is represented by the dark yellow colored boxes, and the far-field is composed by the dark red colored boxes. 

\noindent{\bf Overview of the algorithm}

\begin{figure}
	\centering
%	{\includegraphics[width=150mm]{../figs/tree-5.pdf}}
	{\includegraphics[width=150mm]{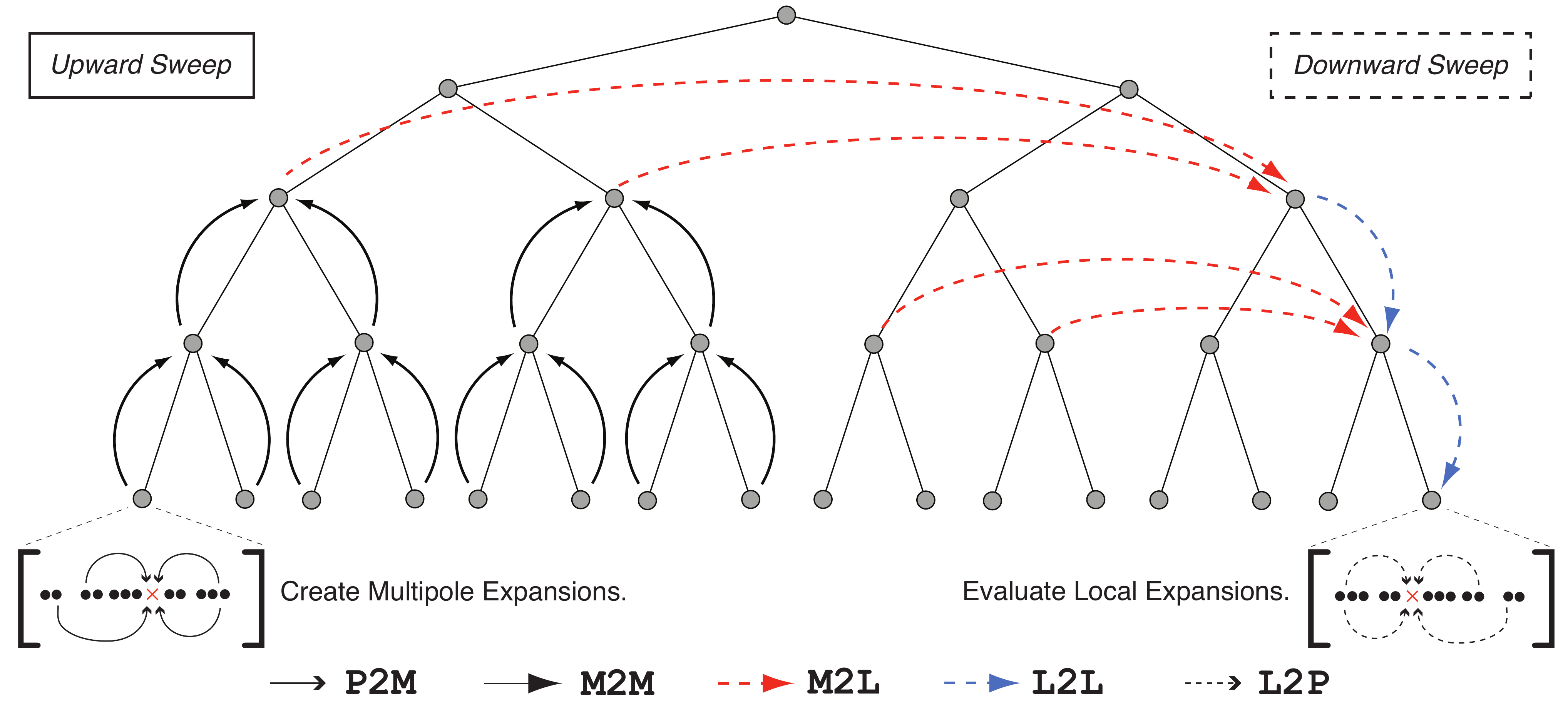}}
	\caption{Overview of the FMM algorithm. The diagram illustrates the \emph{upward sweep} and  the \emph{downward sweep} stages on the tree.  The following operations are illustrated:
\textsc{p}2\textsc{m}--transformation of particles into \ME's (particle-to-multipole);
\textsc{m}2\textsc{m}--translation of \ME's (multipole-to-multipole);
\textsc{m}2\textsc{l}--transformation of an \ME\ into an \LE\ (multipole-to-local);
\textsc{l}2\textsc{l}--translation of an \LE\ (local-to-local);
\textsc{l}2\textsc{p}--evaluation of  \LE's at particle locations (local-to-particle).}
	\label{fig:tree5}
\end{figure}

We use a diagram of the tree structure to illustrate the whole algorithm in one picture, Fig.~\ref{fig:tree5}. The
importance of this presentation is that it relates the control flow and computation to the data structure used by FMM.

After the spatial decomposition stage, the FMM can be summarized in three stages: the upward sweep, the downward sweep, and field evaluation. In the \textit{upward sweep}, \ME's are constructed for each node of the tree. For each leaf node, \ME's  are derived for each particle. On succeeding levels, these expansions are translated to the center of the parent node and combined. This is shown in Fig.~\ref{fig:tree5} by the black arrows going up from the nodes on the left side of the tree. In the \textit{downward sweep} phase, \ME's are first transformed into \LE's for all the cells in the \textit{interaction list} of a given box. This process is represented by the dashed red-colored arrows in Fig.~\ref{fig:tree5}. For a given cell, the interaction list corresponds to the cells of the same level that are not nearest neighbors, but are children of the nearest neighbors of its parent cell. After this series transformation, the \LE's of upper levels are translated to the centers of child cells, and their influence is summed to obtain the complete far-field for each leaf cell. This  process is represented by the dashed blue-colored arrows going down the right side of the tree in Fig.~\ref{fig:tree5}. At the end of the downward sweep, each box will have an \LE that represents the complete far-field for the box. Finally, during the \textit{field evaluation} phase, the total field is evaluated for every particle by adding the near-field and far-field contributions. The near field is obtained by directly computing the interactions between all the particles in the near domain of the box, consisting of nearest neighbor cells in the tree.

In this work, we used the open source \PetFMM package~\citep{CruzKnepleyBarba10} to calculate the fast multipole operation in parallel. The \PetFMM library was designed to offer both high serial performance and scalability, but also to be easily integrated into existing codes. The serial code is completely reused in the parallel setting so that we are never required to maintain two versions of the same algorithm. \PetFMM leverages existing packages to keep its own code base small and clean. Parallel data movement is handled by the \textsl{Sieve} package~\citep{KnepleyKarpeev09} from PETSc~\citep{petsc-user-ref,petsc-web-page}, while load and communication are balanced using a range of different partitioners. In this work we employed either a simple geometric based partitioner which sub-divides the space into $N_x \times N_y \times N_z$ cubes, or the graph partitioner \parmetis~\citep{KarypisKumar98,parmetis-web-page}.
% \hl{Seems like we will be using ptscotch for parallelism. Add references here.}

% ====================================================== %
\section{Numerical experiments} \label{SEC_NumericalExperiments}

\begin{figure*}
\centering
%	\subfigure[Geometry]{\label{FIG_synthetic_model-a}\includegraphics[width=65mm]{../figs/syn_model.pdf}}
	\subfigure[Geometry]{\label{FIG_synthetic_model-a}\includegraphics[width=65mm]{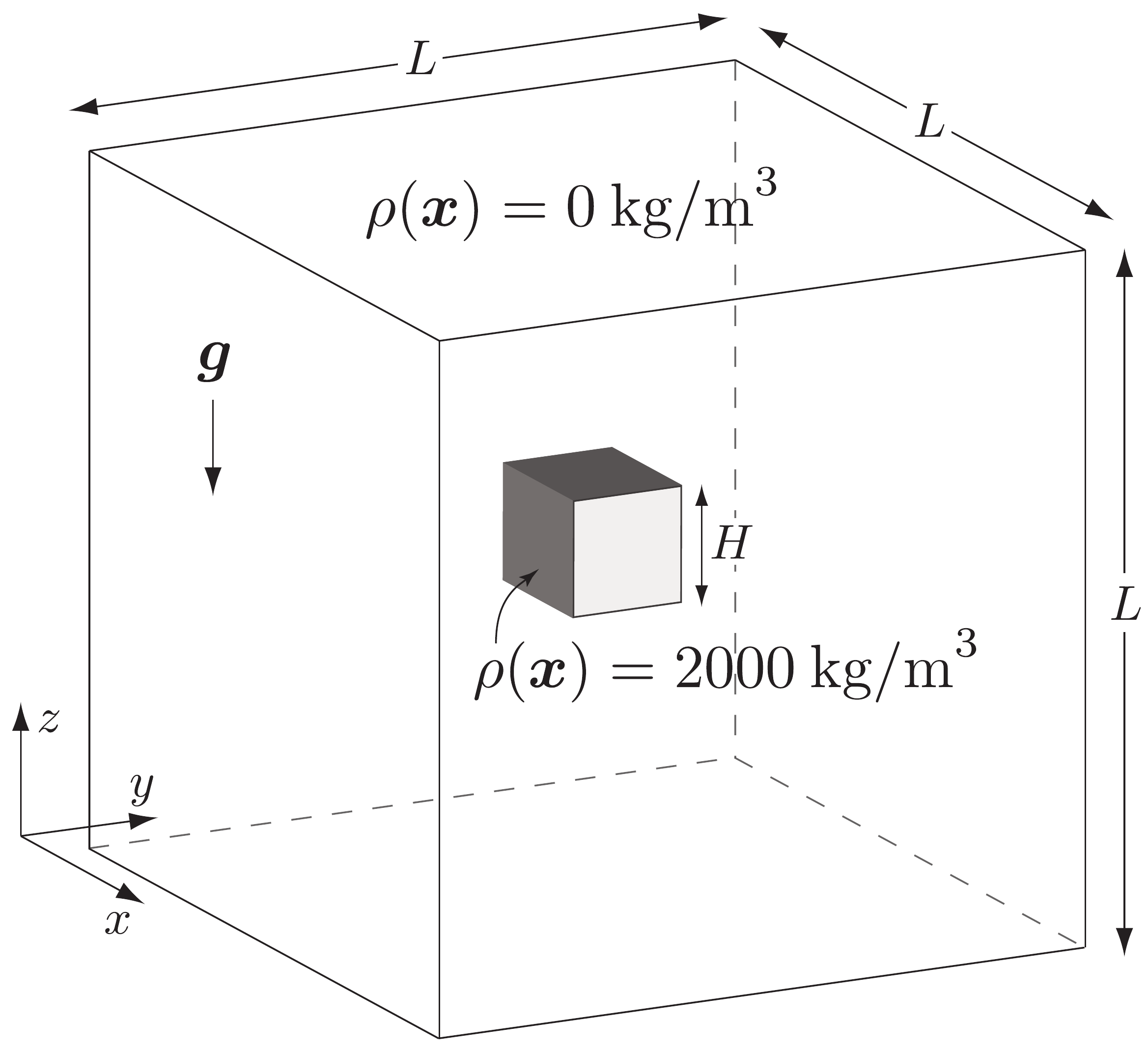}}
%	\subfigure[Gravity field]{\label{FIG_synthetic_model-b}\includegraphics[width=65mm]{../figs/syn_model_gz.pdf}}
	\subfigure[Gravity field]{\label{FIG_synthetic_model-b}\includegraphics[width=85mm]{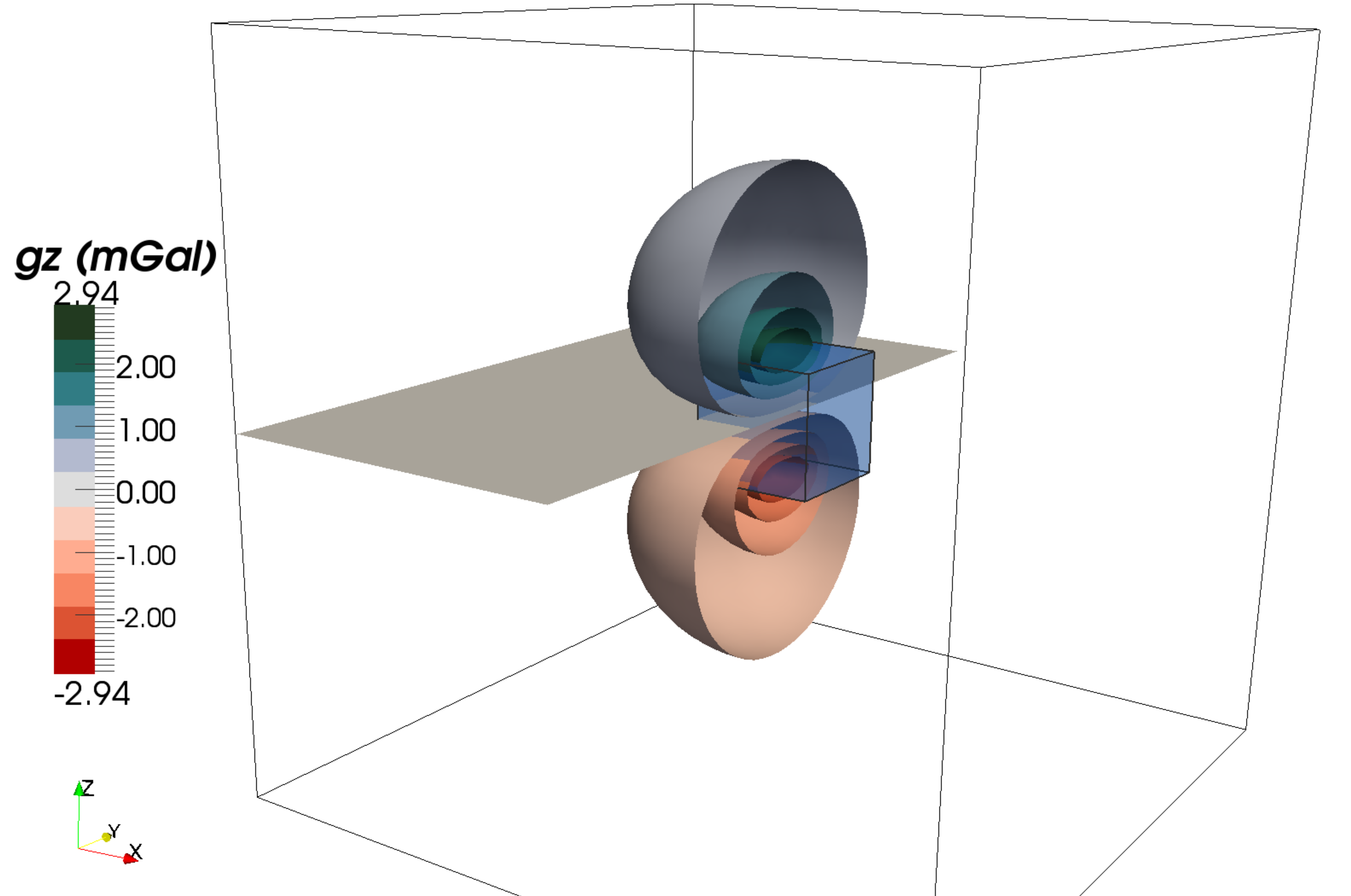}}
	\caption{Synthetic model used thorough out the numerical experiments. 
	(a) Domain and density anomaly and (b) the corresponding analytic gravity field $g_z$ (mGal). 
	The inclusion is indicated by the transparent blue cube.
	See text for dimensions of the domain and density anomaly.}
	\label{FIG_synthetic_model}
\end{figure*}

To understand the discretization error and CPU time required by each of the different classes of forward models, we considered a synthetic gravity model for which we have an analytic solution for the vertical gravity component $g_z$. The model domain $\Omega$ consisted of a cube with side lengths $L = 600$ m, orientated such that $\Omega \equiv [0,600] \times [0,600] \times [-450,150]$ m. Located at the centre of the domain was a cube with side lengths $H = 100$ m, to which we assigned the density, $\rho = 2000$ kg/m$^3$. The surrounding material in the remainder of the domain was regarded as void and assigned a density, $\rho = 0$ kg/m$^3$. The model setup is identical to that used in ~\cite{farq.mosh.417.2009}. By regarding the dense cube as a simple prism, the analytic gravity field can be computed using the closed form expression of ~\cite{li.chou.339.1998}. The model setup and the analytic gravity field component $g_z$ is shown in Fig.~\ref{FIG_synthetic_model}.

	% ====================================================== %
	\subsection{Discretisation error (convergence)} \label{SSEC_DiscretisationError}
The calculations for each numerical method used a mesh comprised of hexahedral elements. 
The number of elements in each direction was chosen such that the density anomaly was exactly resolved by the hexahedral elements. Hence, the error we measure from each method does not include any error due to the discretisation of the density field. We quantify the error in the vertical component of the gravity field $g_z$, using the $L_1$ norm
\begin{equation}
	E_1 = \int_{\Omega} \lvert g_z(\vtr x) - g^h_z(\vtr x) \rvert \, dV,
	\label{EQ_L1norm}
\end{equation}
the $L_2$ norm
\begin{equation}
	E_2=  \left[ \int_{\Omega} \lvert g_z(\vtr x) - g^h_z(\vtr x) \rvert^2 \, dV \right]^{1/2},
	\label{EQ_L2norm}
\end{equation}
and the $L_\infty$ norm
\begin{equation}
	E_\infty = \max_{\vtr x \in \Omega} \medspace \lvert g_z(\vtr x) - g^h_z(\vtr x) \rvert.
	\label{EQ_Linfnorm}
\end{equation}
Here $g_z$ is the exact gravity computed via the analytic solution from ~\cite{li.chou.339.1998}, $g^h_z$ is the approximate gravity field computed using one of three numerical methods (summation, FE, FMM) and $\Omega$ is the model domain.

	% ------------------------------------------------------------------------------------------------------------------------------------------------------- %
	\subsubsection{Summation}

\begin{figure}
\begin{center}
	\includegraphics[width =90mm]{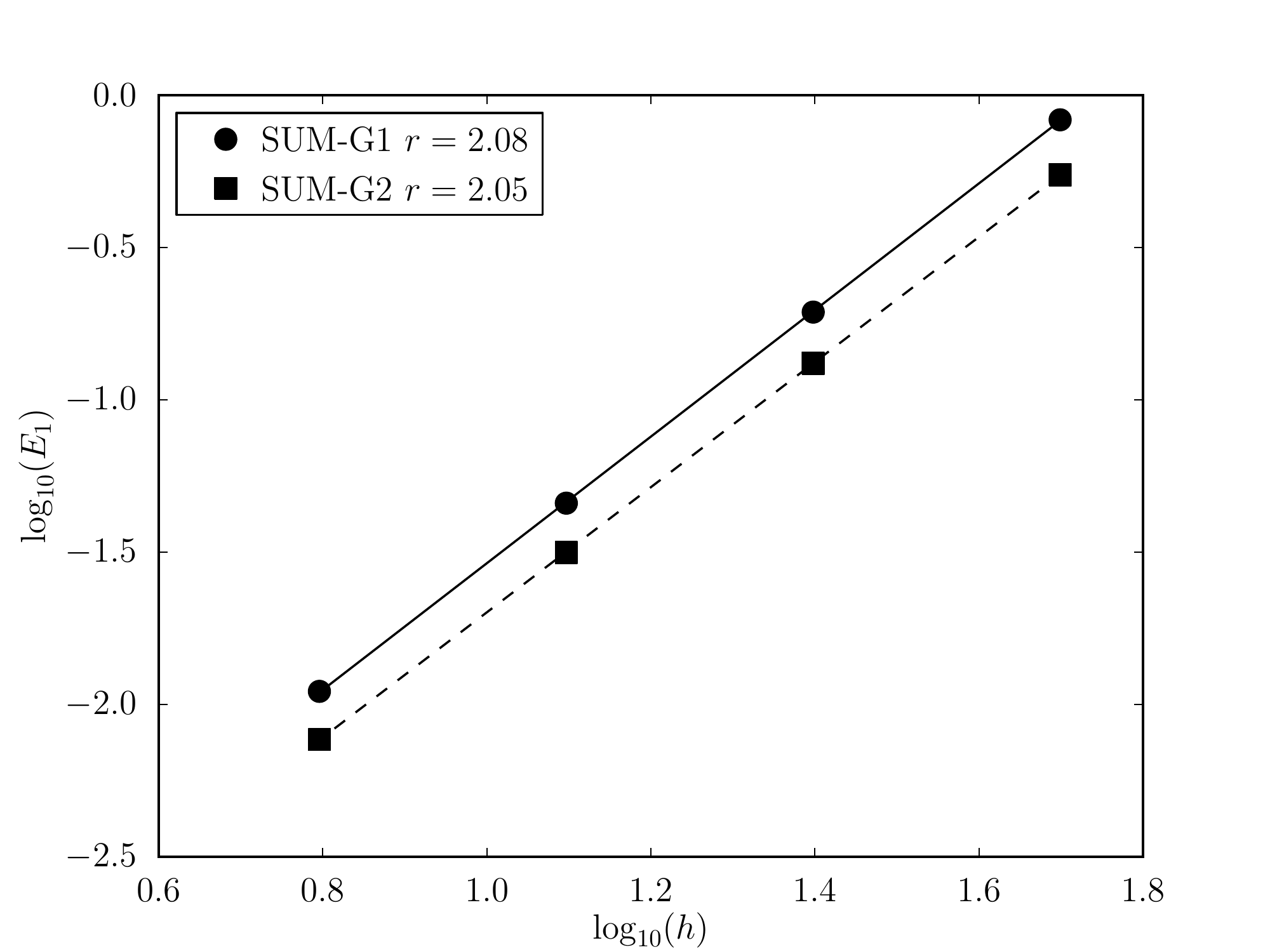} % summation_disc_error.py
	\caption{Convergence rate of the $L_1$ norm for the gravity field computed using \SUMGone and \SUMGtwo.}
	\label{FIG_summation_error}
\end{center}
\end{figure}

We computed the gravity component $g_z$ with \SUMGone and \SUMGtwo using a number of meshes composed of $\bar{M}$ elements in each $x,y,z$ direction. 
The following grid sequence was used to measure the convergence rate, $\bar{M} = \{12, 24, 48, 96 \}$.  The side length of each element is given by $h = 600/\bar{M}$, hence for the mesh sequence used we have $h = \{ 50, 25, 12.5, 6.25 \}$ m.
Given that \SUMAN employs an analytic solution for the gravity at a point due to hexahedral shaped density anomaly, the error expected is of machine precision. Hence, we omit this method from the discussion of errors. The error in Eqs.~\eqref{EQ_L1norm},~\eqref{EQ_L2norm} and ~\eqref{EQ_Linfnorm} was approximated via a 1-point quadrature rule over each hexahedral element in the mesh. The error $E_1$ as a function of grid resolution is shown in Fig.~\ref{FIG_summation_error}. The convergence rate of gravity field in the discrete error measures $E_1, E_2, E_\infty$ is shown in Table~\ref{Table_SUM_convergence_rates}.

\begin{table}
	\caption{Convergence rates obtained with the summation methods.}
	\label{Table_SUM_convergence_rates}
	\begin{tabular}{ccc}
	\hline 
	\bf{error}			&\bf{\SUMGone}	&\bf{\SUMGtwo}		\\
	&&\\
	$E_1$		&2.08	&2.05	\\
	$E_2$		&1.53	&1.52	\\
	$E_\infty$		&0.99	&0.99	\\
	\hline 
	\end{tabular}
\end{table}

	% ------------------------------------------------------------------------------------------------------------------------------------------------------- %
	\subsubsection{Finite element method}

\begin{figure}
\begin{center}
	\includegraphics[width =90mm]{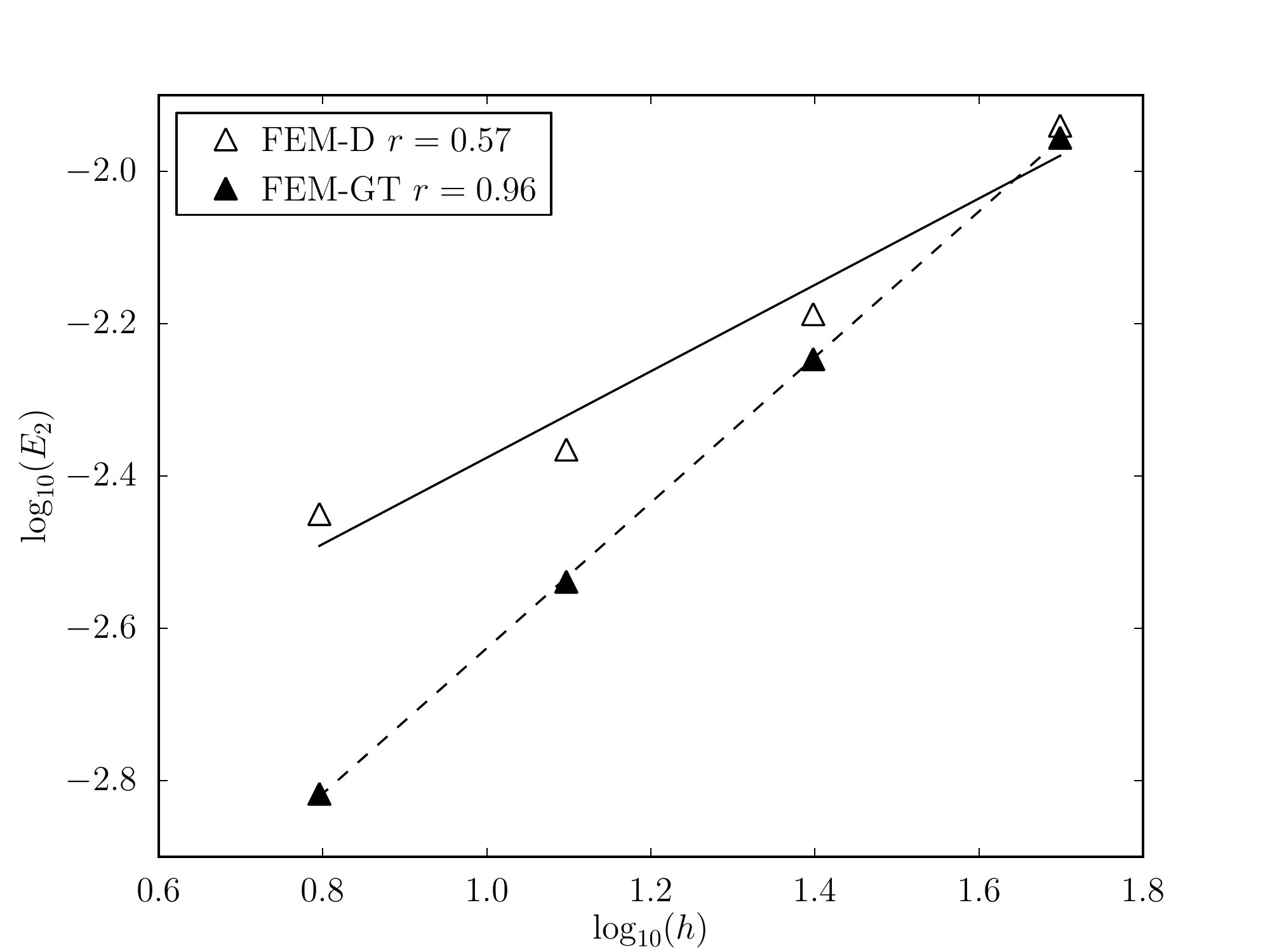} % fem_disc_error.py
	\caption{$L_2$ error of the gravity field computed via \FEMD and \FEMGT.}
	\label{FIG_FEM_error}
\end{center}
\end{figure}

The convergence behavior of the finite element methods \FEMD and \FEMGT was computed using the same grid sequence as in the summation test. Again, the mesh consisted of undeformed elements with $\Delta x = \Delta y = \Delta z = h$. 
A high order Gauss quadrature scheme was used to evaluate the error measures $E_1, E_2, E_\infty$. 
Details of how the error for the FE approaches was computed is provided in Appendix~A.
The $L_2$ discretization error as a function of grid resolution $h$ is shown in Fig.~\ref{FIG_FEM_error}. 
The convergence rate of the gravity field in the discrete $E_1, E_2, E_\infty$ norms is shown in Table~\ref{Table_FEM_convergence_rates}.
From these results, it immediately obvious that using the Robin boundary condition not only produces smaller errors, but the \FEMGT method yields much higher convergence rates.

\begin{table}
	\caption{Convergence rates of the finite element methods for $L/H = 6$.}
	\label{Table_FEM_convergence_rates}
	\begin{tabular}{ccc}
	\hline 
	\bf{error}			&\bf{\FEMD}	&\bf{\FEMGT}		\\
	&& \\
	$E_1$		&0.23	&0.68	\\
	$E_2$		&0.57	&0.96	\\
	$E_ \infty$		&0.97	&0.97	\\
	\hline 
	\end{tabular}
\end{table}

\begin{figure}
\begin{center}
	\includegraphics[width =90mm]{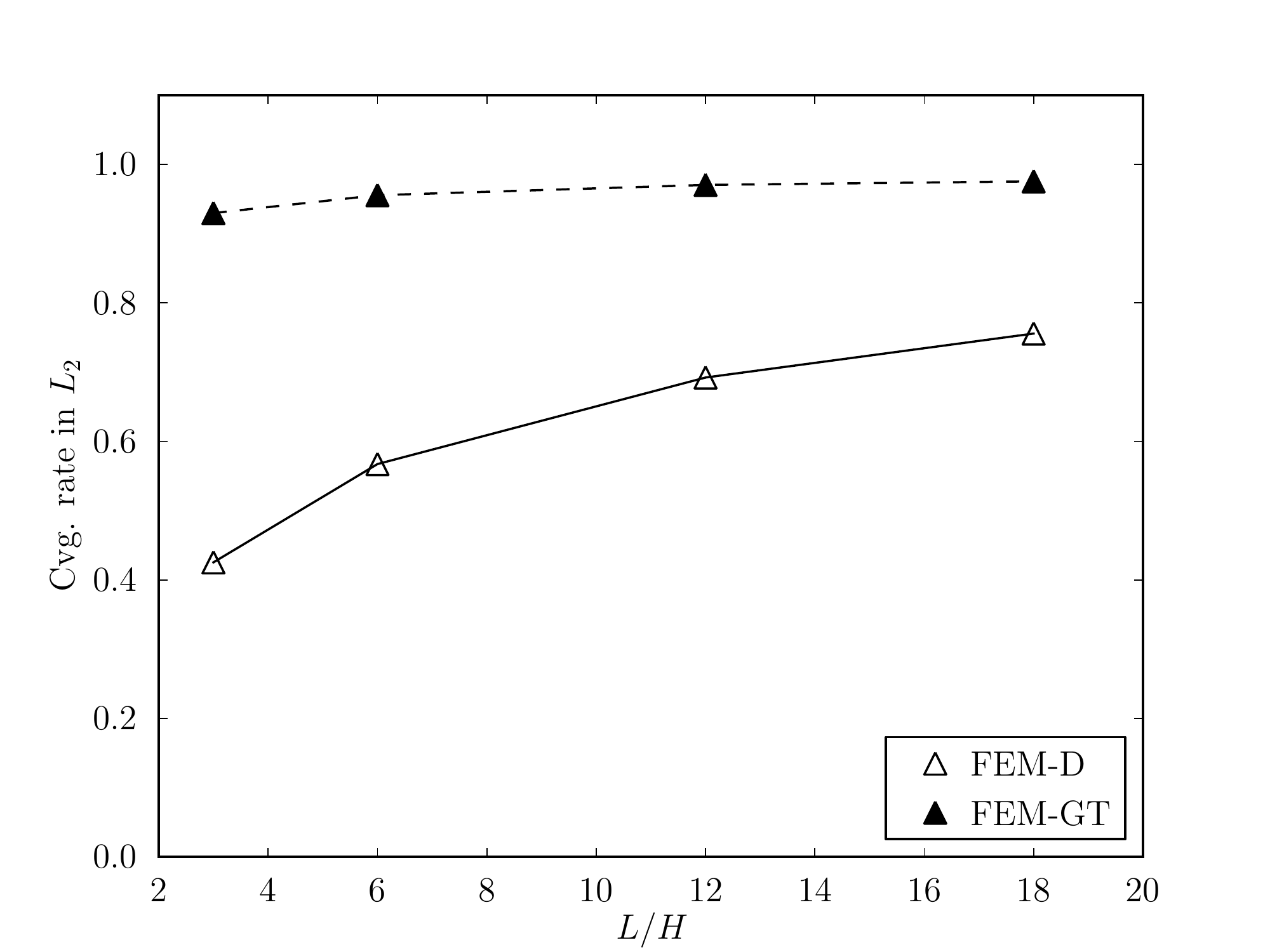} % fem_cvg_domain_data.py
	\caption{Convergence rate in $L_2$ as a function of the domain size.}
	\label{FIG_FEM_convergence}
\end{center}
\end{figure}

To investigate sensitivity of the two boundary conditions used in the FE approaches to the size of the model domain, we performed another convergence test and varied the aspect ratio $L/H$, where model domain and anomaly length are denoted by $L$ and $H$ respectively. 
The anomaly size $H$ was kept fixed at $600$ m, whilst $L$ was increased such that we had the following aspect ratios $L/H  = \{ 3, 12, 18 \}$. As in the other convergence tests, four meshes of increasing resolution were used. To keep the discretization errors comparable between the different models, we ensured that element size on each of the four meshes, for each $L/H$ yielded element sizes of $h = \{ 50, 25, 12.5, 6.25 \}$ m. The $L_2$ convergence rates are shown in Fig.~\ref{FIG_FEM_convergence}. Here we see that the convergence rate of \FEMGT is independent of the domain size, whilst the convergence rate of the gravity field computed using \FEMD increases as the model domain increases. We expect that the rate from \FEMD approaches 1.0 as $L/H \rightarrow \infty$. 

	% ------------------------------------------------------------------------------------------------------------------------------------------------------- %
	\subsubsection{Fast multipole method}

%\begin{figure}
%\begin{center}
%	\includegraphics[width =80mm]{../figs/fmm_convergence_dx.pdf} % fmm_linear_reg_cvg.py
%	\caption{Convergence rate of the gravity field computed via \FMM. Here $p=8$.
%	\label{FIG_FMM_error}}
%\end{center}
%\end{figure}

\begin{figure}
\begin{center}
	\includegraphics[width =90mm]{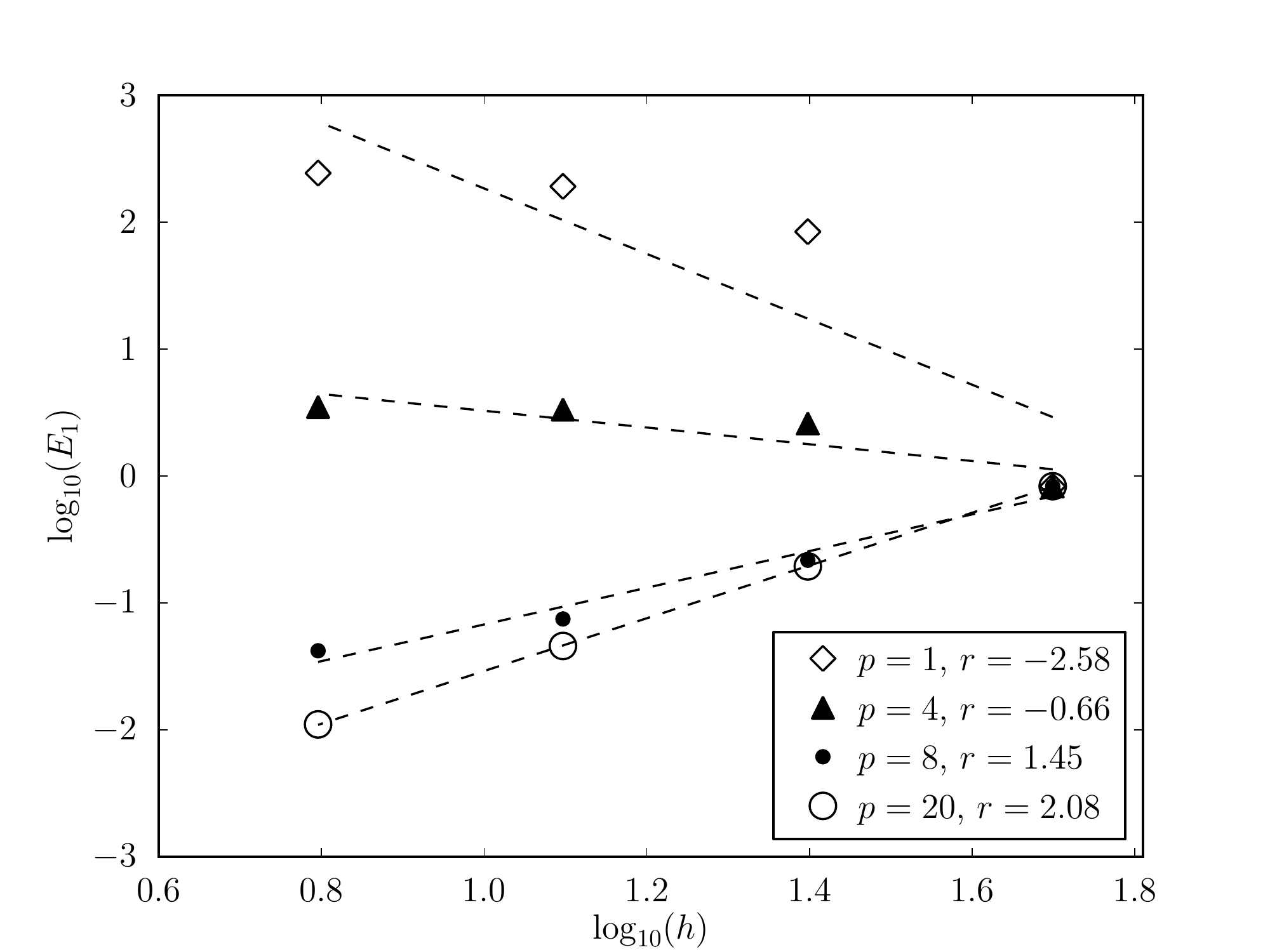} % fmm_disc_error.py
	\caption{Convergence rate of the $L_1$ norm of the gravity field computed using \PetFMM using different values of $p$.}
	\label{FIG_FMM_p_error}
\end{center}
\end{figure}

The convergence rate of \PetFMM was performed using the same mesh sequence as in the summation experiments. As for the summation methods, the error measures were approximated via a 1-point quadrature rule over each hexahedral element.  The accuracy of the solution obtained via \PetFMM is strongly related to the number of terms $p$ used in the expansion of Eq.~\eqref{eq:decouple}. The measured convergence rate in the different norms are presented for $p = \{ 1,4,8,20\}$ in Table~\ref{Table_FMM_convergence_rates}. For the error measure $E_1$, we show the variation with grid resolution $h$ in Fig.~\ref{FIG_FMM_p_error}. Comparing with the rates from the summation methods from Table~\ref{Table_SUM_convergence_rates}, we note that as $p$ increases, the convergence rates of \PetFMM approach those obtained using \SUMGone and \SUMGtwo.
\begin{table}
	\caption{Convergence rates of \PetFMM using different values of $p$.}
	\label{Table_FMM_convergence_rates}
	\begin{tabular}{c c cccc}
	\hline 
	\bf{error}			&&\multicolumn{4}{c}{$\boldsymbol p$} \\
				&&\bf{1}		&\bf{4}		&\bf{8}	&\bf{20}		\\
%	\cline{1-1} 
%	\cline{3-6} 
			&&		&		&	&		\\
	$E_1$		&&-2.58	&-0.66	&1.45	&2.08	\\
	$E_2$		&&-1.81	&0.08	&1.51	&1.53	\\
	$E_ \infty$	&&-1.76	&-0.04	&0.99	&0.99	\\
	\hline 
	\end{tabular}
\end{table}

	% ====================================================== %
	\subsection{Optimality (CPU time)} \label{SSEC_Optimality}
	Here we report the CPU time of the different numerical methods applied to the synthetic model described in Sec.~\ref{SEC_NumericalExperiments}.
All timings reported were obtained with code compiled using GCC 4.4.3 with level three optimization and with an optimized build of the PETSc library. The timing runs were performed on {\it Octopus}, which is an 8-core Intel Xeon 2.67GHz (Nehalem) machine possessing 64 GBytes of RAM.
	
	% ------------------------------------------------------------------------------------------------------------------------------------------------------- %
	\subsubsection{Summation}
On a given mesh, the time required for the summation methods is proportional to the number of locations where the gravity is evaluated.  For this series of tests, we evaluated the gravity on a regularly spaced array of $150 \times 150$ points, located at the upper surface of the model domain.
In Table~\ref{Table_SUM_timings}, we report the total CPU time (sec)  per gravity station on the following sequence of meshes, $\bar{M} = \{ 6,12,24,48,96 \}$. 
Methods \SUMGone and \SUMGtwo compute the three components of the gravity vector, whilst \SUMAN and  \SUMGone$\!\!(z)$ only compute the gravity field in the $z$ direction. 
\begin{table}
	\caption{CPU time (sec) for the summation methods. The times reported are normalized by the number of locations where the gravity field was evaluated. Here $\bar{M}$ is the number of cells used to discretize the subsurface in each direction and $h$ is side length (m) of each cell.}
	\label{Table_SUM_timings}
	\begin{tabular}{cc c cccc}
	\hline 
			&			&	&\multicolumn{4}{c}{\bf{CPU time (sec) / station}} \\
	$\boldsymbol h$ \bf{(m)}	&$\boldsymbol{\bar{M}}$	&	&\bf{\SUMGone}$\boldsymbol{(z)}$	&\bf{\SUMGone}	&\bf{\SUMGtwo}	&\bf{\SUMAN} \\ 
%	\cline{1-2} 
%	\cline{4-7}
			&			&	&				&			&			& \\
	%
	%\vspace{-1.8mm} \\
	100	&6		&&4.78e-07	&7.36e-07	&5.34e-06	&7.39e-05 \\
	50	&12		&&3.71e-06	&5.72e-06	&4.29e-05	&5.50e-04 \\
	25	&24		&&2.98e-05	&4.56e-05	&3.40e-04	&4.39e-03 \\
	12.5	&48		&&2.37e-04	&3.65e-04	&2.68e-03	&3.76e-02 \\
	6.25	&96		&&1.90e-03	&2.92e-03	&2.13e-02	&2.83e-01 \\
	\hline 
	\end{tabular}
\end{table}
All methods possess an approximately linear relationship between the CPU time / station and the number of cells used to discretize the domain.  Considering the one point quadrature rule methods, \SUMGone is only a factor of 1.5 slower than \SUMGone$\!\!(z)$. The slight increase in time required for \SUMGone is a consequence of a more general quadrature. In this implementation, arbitrarily deformed hexahedral elements are permitted, whilst element edges were required to be perpendicular to the coordinate system in \SUMGone$\!\!(z)$. Allowing the elements to be deformed requires that the integration be performed in a reference coordinate system, which thus requires the inverse Jacobian (coordinate transformation) to be evaluated. 
\SUMGtwo was observed to be approximately 7 times slower than \SUMGone, even though it employs 8 times as many quadrature points.
%\SUMGtwo was expected to be roughly 8 times slower than \SUMGone as it employs $2 \times 2 \times 2$ quadrature points compared to a single quadrature point. 
%However, we observe that it is only $\sim 7$ times slower. 
%We attribute this to ...? 
The closed form method, \SUMAN is $\sim150$ times slower than \SUMGone$\!\!(z)$ and $\sim 13$ times slower than \SUMGtwo.

	% ------------------------------------------------------------------------------------------------------------------------------------------------------- %
	\subsubsection{Finite element method}
	The FE calculations were performed using meshes consisting of $\bar M = \{ 12, 24, 48, 96, 192, 384 \}$ elements in each direction. 
%The gravity field was evaluated at the centroid of each cell. 
In all the calculations performed, the multigrid preconditioner used a coarse grid consisting of $6 \times 6 \times 6$ elements. The number of grid levels $n_l$, was chosen to give the desired value of $\bar M$ on the finest grid level.
%A grid refinement factor of two was used between all grid levels. 
	
In Table~\ref{Table_FEM_timings} we report the time required to perform the linear solve of the system in Eq.~\eqref{EQ_discretepoisson} using both \FEMD and \FEMGT. The time required for the solve represented more than 99\% of the total execution time, thus only the solve time is reported. We observe that the number of iterations required by both methods are independent of the grid resolution. Furthermore, both the CPU time and memory usage scale approximately linearly with respect to the number of unknowns in the potential field, $n = (\bar{M}+1)^3$. The solve time for the \FEMD method is slightly higher than that required by \FEMGT. The difference in CPU time is attributed to the manner in which the Dirichlet boundary conditions were imposed during each application of the matrix free product, $\mtx A \vtr y$. This particular operation could easily be further optimized in the future.
\begin{table}
	\caption{Performance of the FE methods. The CPU time (sec) and the number of iterations required by the Poisson solver are reported. The memory usage (MB) for \FEMD and \FEMGT are the same and are reported in the final column.}
	\label{Table_FEM_timings}
	\begin{tabular}{cc c cc c cc c c}
	\hline 
			&			&	&\multicolumn{2}{c}{\bf{\FEMD}}	& &\multicolumn{2}{c}{\bf{\FEMGT}}  \\
	$\boldsymbol h$ \bf{(m)}	&$\boldsymbol{\bar{M}}$	&	&\bf{CPU time (sec)}	&\bf{Iter.}
						&	&\bf{CPU time (sec)}	&\bf{Iter.} 	
						&	&\bf{Mem. (MB)} 	 \\ 
%	\cline{1-2} 
%	\cline{4-5}
%	\cline{7-8}
%	\cline{10-10}
		&		&&			&		&&			&			&& \\
	%
	%\vspace{-1.8mm} \\
	50	&12		&&1.17e-02	&8		&&1.07e-01	&8			&&$< 10$	\\
	25	&24		&&1.10e+00	&9		&&1.01e+00	&9			&&$< 10$\\
	12.5	&48		&&8.87e+00	&9		&&8.08e+00	&9			&&4.00e+01\\
	6.25	&96		&&7.11e+01	&9		&&6.45e+01	&9			&&2.85e+02\\
	3.13	&192	&&5.67e+02	&9		&&4.68e+02	&8			&&2.20e+03\\
	1.56	&384	&&4.15e+03	&8		&&3.74e+03	&8			&&1.70e+04\\
	\hline 
	\end{tabular}
\end{table}

	% ------------------------------------------------------------------------------------------------------------------------------------------------------- %
	\subsubsection{Fast multipole method}
	The performance of the \PetFMM algorithm was measured using the same sequence of meshes as used in the FE approaches, i.e. the mesh contained $\bar M = \{ 12, 24, 48, 96, 192, 384 \}$ elements in each direction. The octree used to define the FMM data structure used $k = 2^L$ cells along each axis, where $L$ denotes the number of levels within the tree. For the mesh sequence used, we employed $L = \{ 2, 3, 4, 5, 6, 7 \}$. In these calculation presented, the gravity vector was computed at the centroid of each cell used to discretize the density field. In Table~\ref{Table_FMM_timings} we report the CPU time (sec) required to execute the \PetFMM algorithm. The time required to evaluate the gravity field is negligible compared to time spend in the \PetFMM algorithm and is thus not reported here. For these experiments, the graph partitioner \parmetis was used.
	
\begin{figure}
\begin{center}
	\includegraphics[width =90mm]{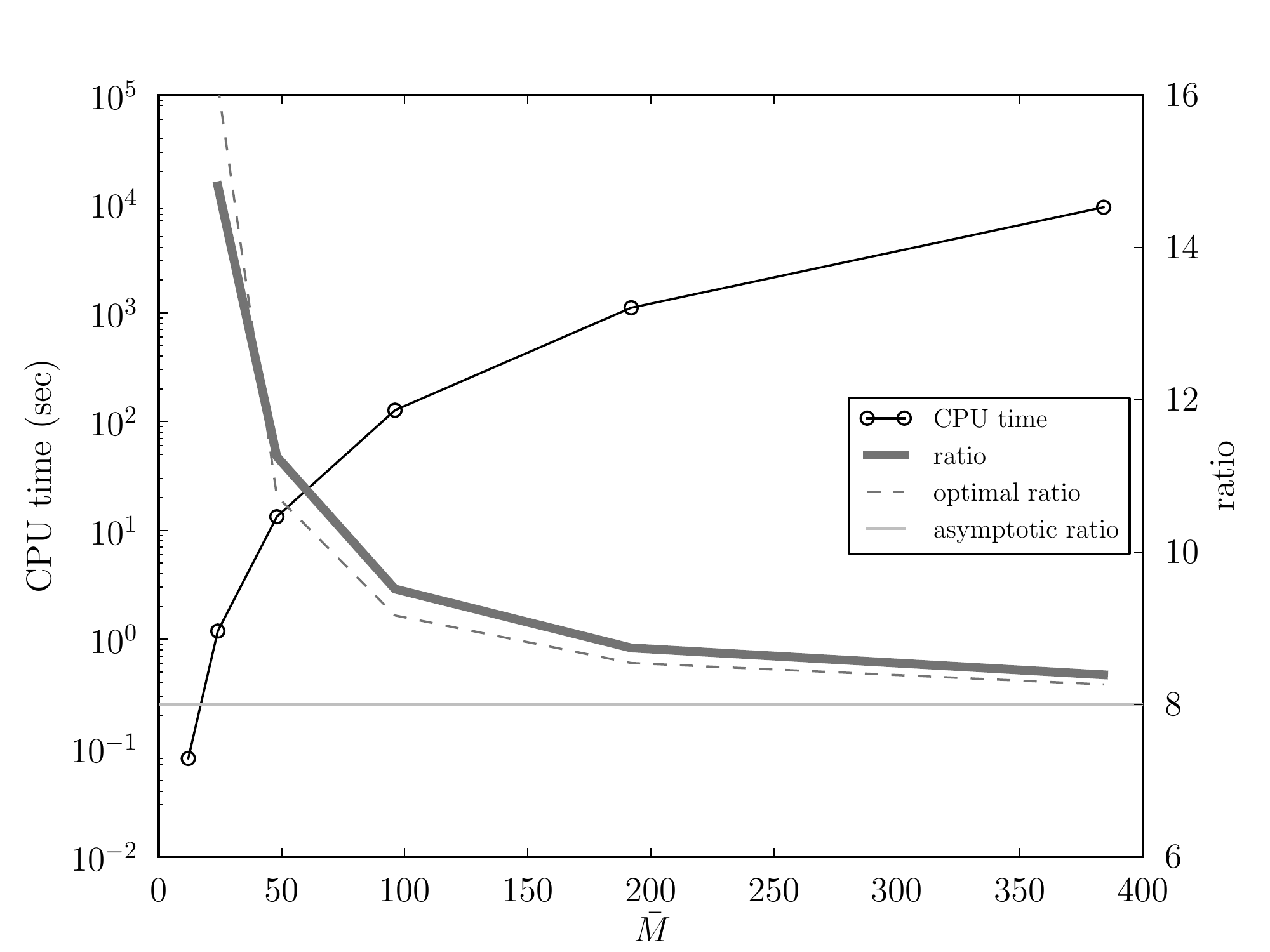}  % fmm_serial_time_ratio5.py
	\caption{Computation time as a function of the number of density blobs $\bar{M}^3$ for \PetFMM. The left $y$-axis denotes CPU time (sec) and right $y$-axis denotes the ratio of solution times between the current and previous grid resolution. For the grid sequence used, the asymptotic (linear) scaling would yield a ratio of 8, here denoted via the dashed line.}
	\label{FIG_FMM_optimal}
\end{center}
\end{figure}

	For the sequence of meshes used in our test, an optimal FMM algorithm may be expected to yield execution times and memory usage requirements which increased by a factor of eight, for each increase in grid resolution. From Table~\ref{Table_FMM_timings}, the memory usage is observed to follow this scaling. However, we note that the CPU time for \PetFMM is observed to only approach the anticipated result as $\bar{M}$ increases. In Fig.~\ref{FIG_FMM_optimal} the solution time (solid thin line, left $y$-axis) and the solution time ratio for $t_{k}/t_{k-1}$ (solid thick line, right $y$-axis), is plotted as a function of the number of elements in each direction $\bar{M}$. The anticipated optimal value of $t_{k}/t_{k-1} = 8$ is denoted via the thin gray line. 
%We note that the anticipated optimal value is approached asymptotically as more voxels are used. 

	We can explain the deviation of this ratio observed with small numbers of voxels to a surface to volume effect. For
a cube, divided into $k$ pieces along each axis, we obviously have $k^3$ small constituent cubes. Of these, 8 are corner
cubes which have 7 neighbors. There are 12 edges of the large cube, each of which has $k-2$ small cubes with 11
neighbors. Similarly, there are 6 faces of the large cube, each of which has $(k-2)^2$ small cubes with 17
neighbors. The remaining $(k-2)^3$ interior cubes have 26 neighbors. We can check that the number of small cubes is
correct,
\begin{eqnarray}
  & & (k-2)^3 + 6 (k-2)^2 + 12 (k-2) + 8\\
  &=& (k^3 - 6k^2 + 12k - 8) + 6 (k^2 - 4k + 4) + 12 (k-2) + 8\\
  &=& k^3.
\end{eqnarray}
If we assume that $B$ particles are in every cube, then the direct work done per cube is given by
\begin{equation}
  W_c = \frac{B (B-1)}{2} + N_B B^2 \approx \left(N_B + \frac{1}{2} \right) B^2,
\end{equation}
where $N_B$ is the number of cube neighbors. The ratio of work $R$, between a $2k$ division compared to a $k$
division along each axis is given by,
\begin{eqnarray}
  R \left( \frac{2k}{k} \right) &=& \frac{(2k-2)^3 \frac{53}{2} + 6 (2k-2)^2 \frac{35}{2} + 12 (2k-2) \frac{23}{2} + 8 \frac{15}{2}}{(k-2)^3 \frac{53}{2} + 6 (k-2)^2 \frac{35}{2} + 12 (k-2) \frac{23}{2} + 8 \frac{15}{2}} \\
    &=& \frac{53 (2k-2)^3 + 210 (2k-2)^2 + 276 (2k-2) + 120}{53 (k-2)^3 + 210 (k-2)^2 + 276 (k-2) + 120}   \\
    &=& \frac{(8k^3 - 24k^2 + 24k - 8) + 3.96 (4k^2 - 8k + 4) + 5.21 (2k-2) + 2.26}{(k^3 - 6k^2 + 12k - 8) + 3.96 (k^2 -4k + 4) + 5.21 (k-2) + 2.26} \\
    &=& \frac{8 k^3 - 8.16 k^2 + 2.74 k - 0.32}{k^3 - 2.04 k^2 + 1.37 k - 0.32}. \\
    \label{eq:fmm_optimal_work_scaling}
\end{eqnarray}
In the first two tests considered in Table \ref{Table_FMM_timings}, we have $k = 2$ and $4$, thus
\begin{equation}
  R\left( \frac{4}{2} \right) = \frac{968.00}{60.00} = 16.13.
\end{equation}
%Note that we undershoot this number due to the completely scalable work done in the \textsc{m}2\textsc{l}--transformation. 
Even at $k = 8$ we have
\begin{equation}
  R \left( \frac{16}{8} \right) = \frac{95288.00}{10392.00} = 9.17
\end{equation}
and we can see that not inconsiderable surface-to-volume effects persist for larger octrees. 
The optimal ratio defined by Eq. \eqref{eq:fmm_optimal_work_scaling} is denoted in Fig.~\ref{FIG_FMM_optimal} via the dashed line.
%Given these ratios, the curve given in Fig.~\ref{FIG_FMM_optimal} is optimal.
The agreement between the optimal and measure work scaling illustrated in Fig.~\ref{FIG_FMM_optimal} verify the optimality of the  \textsc{m}2\textsc{l}--transformation.

\begin{table}
	\caption{CPU time (sec) and memory usage (MB) for \PetFMM with increasing grid resolution. In these calculations we used an expansion order of $p=8$. We note that the memory counter used in the implementation of \PetFMM was not able to represent the number of bytes required for the case $\bar{M}=384$. }
	\label{Table_FMM_timings}
	\begin{tabular}{ccc c cc}
	\hline 
%	\vspace{-3.0mm} \\
	$\boldsymbol h$ \bf{(m)}	&$\boldsymbol{\bar{M}}$ &$\boldsymbol L$	&	&\bf{CPU time (sec)} 	&\bf{Mem. (MB)} \\ 
%	\cline{1-3} 
%	\cline{5-6}
		&	&		&&			& \\
	%
%	\vspace{-1.8mm} \\
	50	&12	&2		&&8.02e-02	&$<$ 1.00e+00	 \\
	25	&24	&3		&&1.19e+00	&1.67e+00	 \\
	12.5	&48	&4		&&1.34e+01	&1.34e+01	 \\
	6.25	&96	&5		&&1.27e+02	&1.07e+02	 \\
	3.13 &192		&6	&&1.11e+03	&8.56e+02	\\
	1.56 &384		&7		&&9.33e+03	&(counter overflow) \\ %&\hl{2.55e+03 ??? Not x 8!!! }  \\
	\hline 
	\end{tabular}
\end{table}

	% ====================================================== %
	\subsection{Parallel scalability} \label{SSEC_Scalability}
In order to measure the parallel performance of an algorithm, two types of studies are typically employed. The first measure considers {\it weak scaling}, in which a fixed number of unknowns per processor (i.e the work per processor) is kept constant and more processors are introduced. Thus the overall problem size increases with the number of processors, but the work per process remains constant.  Ideal weak scaling would yield a solution time which was independent of the number of processors which were employed. Alternatively, {\it strong scaling} considers a problem with a fixed number of unknowns which is solved using an increasing number of processors. Thus, the unknowns per processor decreases as the number of processors increases. Ideal strong scaling would yield a solution time which linearly decreases in proportion to the number of processors used to solve the problem. 

In the interest of developing fast algorithms for performing gravity inversions in a reduced amount of time, here we only consider the strong scalability of the three algorithms presented. If a simulation required $t_0$ seconds on $p_1$ processors, the optimal time $t_{opt}$, on $p_2 > p_1$ processors is $t_{opt} = t_0 (p_1/p_2)$. The parallel efficiency $E$ of the strong scaling is measured according to
\begin{equation}
	E = 100 \left( \frac{t_{opt} }{ t_{measured}} \right)	,
\end{equation}
where $t_{measured}$ is the measured time taken for the computation on $p_2$ processors.
All parallel results presented here were performed on the CADMOS IBM Blue Gene/P (\url{http://bluegene.epfl.ch}). 

	% ------------------------------------------------------------------------------------------------------------------------------------------------------- %
	\subsubsection{Summation}

All of the summation algorithms considered here exploit parallelism by sub dividing the set of voxels used to represent the density structure amongst $n_p$ processors. The spatial decomposition of the mesh was defined by slicing the domain into $N_x, N_y, N_z$ subdomains such that $n_p = N_x \times N_y \times N_z$. The only communication required in our implementation is the global reduction (sum) of a vector of length equal to the number of evaluation points. Thus if the number of voxels in each processors subdomain is equal, the only departure from perfect strong scaling can be attributed to the single call to {\tt MPI\_Allreduce}. In Table~\ref{Table_SUM_strongscaling} we report the CPU times obtained from using \SUMGone$\!\!(z)$ with a model domain of $128^3$ voxels and $100^2$ evaluation points which were regularly spaced in a horizontal plane located at the upper surface of the model. 
Both the CPU time for the total computation and the time for the global reduction are reported. We note the time for the global reduction does not exhibit perfect strong scaling for this set of experiments. Accordingly, when the time required to perform the evaluation and local sum of the gravity contributions is much larger than the time required for the reduction, excellent scalability is observed ($n_p \le 256$). When this time is comparable with the cost of the reduction, the sub optimal scaling of the reduction will become significant and deteoriate the scaling of the total execution time. Comparing the total CPU times for $n_p=8$ and 2048, we observe a parallel efficiency of $E \approx 78 \%$.
\begin{table}
	\caption{Strong scaling for \SUMGone on CADMOS BG/P, using a mesh with $128^3$ cells and $100^2$ evaluation points. Here $n_p$ indicates the number of processors used. (D) indicates the job was executed in DUAL mode, implying two processors per node were used. (V) indicates the job was launched in VN mode, in which all four processors per node were used.}
	\label{Table_SUM_strongscaling}
	\begin{tabular}{l c cc}
	\hline 
				&	&\multicolumn{2}{c}{\bf{CPU time (sec)}} \\
	$\boldsymbol{n_p}$		&	&\bf{Total} 	&\bf{Reduction}	 \\ 
%	\cline{1-1} 
%	\cline{3-4}
			&&				&	\\
	%
	%\vspace{-1.8mm} \\
1      			 &&1.0632e+03     &1.6999e-04 \\
8   			 &&1.3922e+02     &1.2458e+01 \\
64			 &&1.8222e+01      &3.1192e+00 \\
128   		&&9.3883e+00      &2.0806e+00 \\
$256$ (D)     	&&4.8376e+00      &1.3013e+00  \\
$512$ (V)    	&&2.4939e+00      &7.8258e-01  \\
512     		&&2.4934e+00      &7.8200e-01 \\
$2048$ (V)   	&&7.0128e-01      &3.2787e-01 \\
	\hline 
	\end{tabular}
\end{table}

	% ------------------------------------------------------------------------------------------------------------------------------------------------------- %
	\subsubsection{Finite element method}
The success of a parallel multigrid is largely dependent on the type of coarse grid solver used. 
We consider a direct extension of sequential multigrid algorithms which employ a direct solver on the coarsest grid level. 
The direct solve on the coarse grid was performed in parallel using either the  multi-frontal method MUMPS \citep{ame.etal.41.2001}, or by TFS \citep{tufo.fisc.151.2001}. MUMPS is a general purpose parallel direct solver, whilst TFS is specifically designed for matrix problems in which a processors subdomain contains very few degrees of freedom (as is the case on our distributed coarse grid). TFS has the limitation that the number of processors must be a power of two.

\begin{figure}
\begin{center}
	\includegraphics[width =110mm]{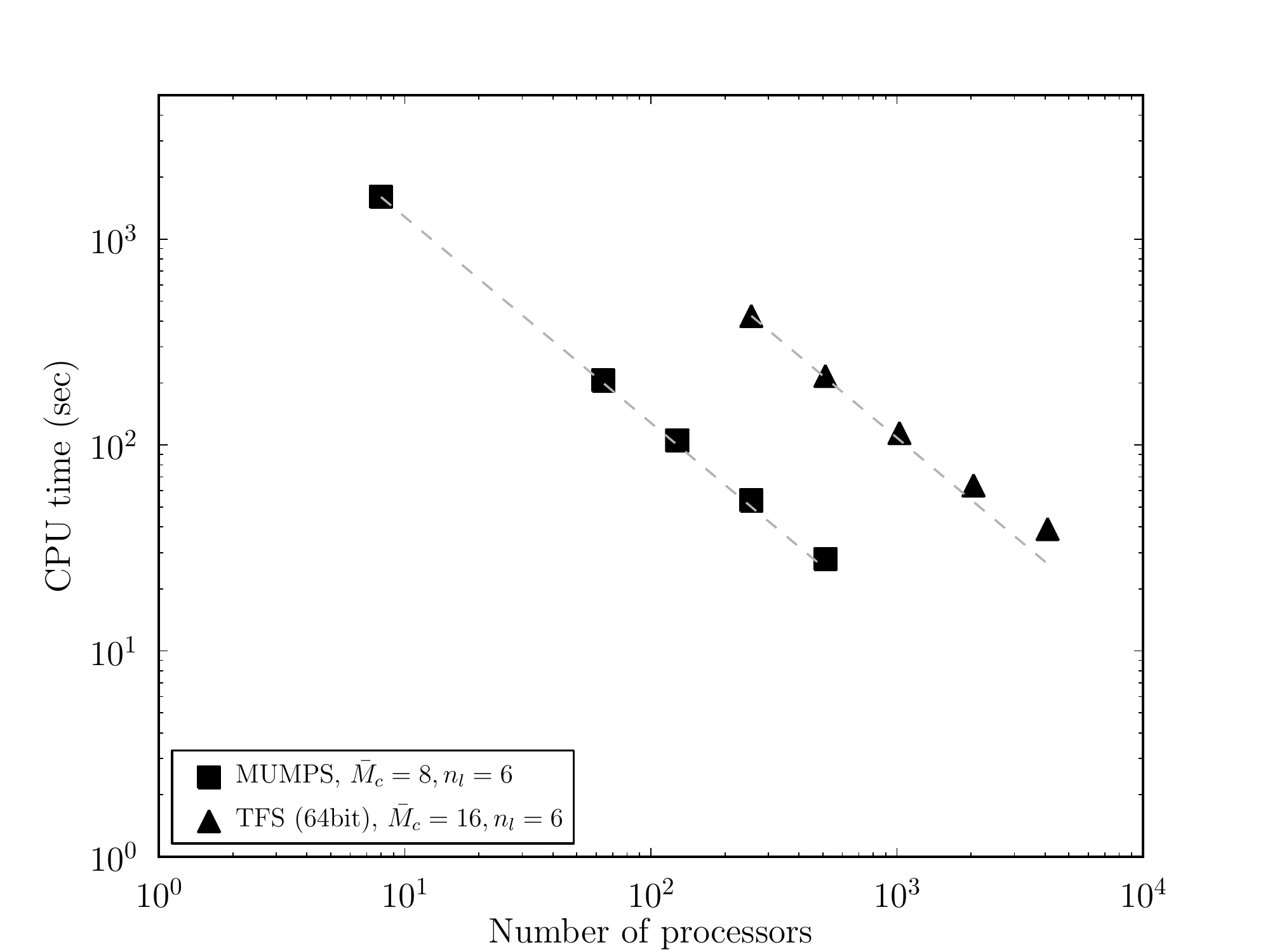} % fem_strongscaling.py 
	\caption{Strong scaling on the CADMOS BG/P for two different resolution \FEMD simulations. The optimal time is indicated by the dashed line. The coarse grid consisted of $\bar{M}_c$ elements in each direction and each model used $n_l=6$ levels. Two different coarse grid solvers, MUMPS and TFS were employed (see text for further details).}
	\label{FIG_FEM_strongscaling}
\end{center}
\end{figure}

To examine the strong scalability, we considered two experiments in which the fine grid contained either $256^3$ elements of $512^3$ elements. The coarse grid was defined via $\bar{M}_c$ elements in each direction. Both experiments used six grid levels, with $\bar{M}_c$ being 8 and 16 respectively. 
In our geometric multigrid implementation, we require for a given grid, that each processor's local subdomain must contain at least one element. 
Accordingly, the number of elements in the coarse grid thus places an upper limit on the maximum number of CPU's we can use.
The results of the strong scalability are shown in Fig.~\ref{FIG_FEM_strongscaling}. The scalability of \FEMD and \FEMGT are expected to be identical so only the results of \FEMD are presented. The measured parallel efficiency on 512 CPUs was $E \approx 90\%$ for the problem using MUMPS and $E \approx 68\%$ on 2048 CPUs for the problem employing TFS.

	% ------------------------------------------------------------------------------------------------------------------------------------------------------- %
	\subsubsection{Fast multipole method}
    To examine the strong scalability of \PetFMM, we considered three different meshes with $\bar{M}^3$ elements where
    $\bar{M} = \{ 96, 192, 384 \}$. For a given number of input density values, there is a number of levels $L_v$ which
    minimizes the total computation time.	As in the multigrid implementations, they are certain restrictions upon the
    number of CPU's $(n_p)$ which can be used with \PetFMM. The primary constraint is on the number of local trees in
    the spatial decomposition. The number of local trees $N_t$ is given by $2^{d \times r_l}$, where $d=3$ is the
    spatial dimension and $r_l$ is the root level of the tree. For efficiency, it is required that $N_t > n_p$, so that
    at least one tree is distributed to every process.

For the parallel runs presented here, the simple geometric based partitioning algorithm was used to balance load and communication.
The total execution times are reported in Table~\ref{Table_FMM_strongscaling}. 
	
\begin{figure}
\begin{center}
	\includegraphics[width =130mm]{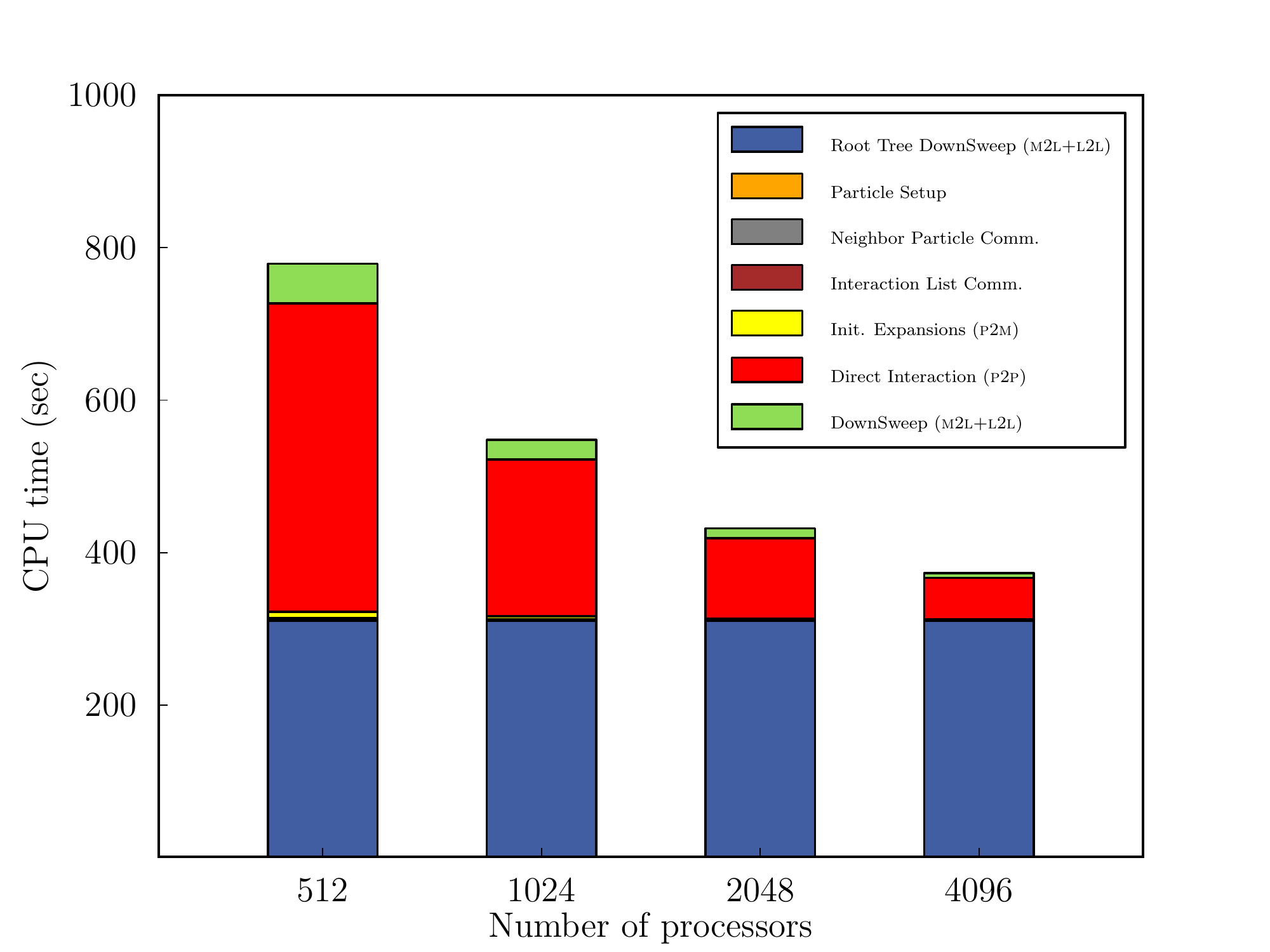} % fmm_strong_scaling_bar2.py 
	\caption{Breakdown of the strong scalability of the individual \PetFMM components.
	Note that not all components listed in the legend are visible in the bar chart as they represent a very small fraction of the total execution time.}
	\label{FIG_FMM_strongscaling}
\end{center}
\end{figure}

The strong scaling efficiency is observed to decrease as the number of processors used increases and also as the root level increases. To better understand the reason for this scaling behavior, we examined the scalability of individual components within the \PetFMM implementation. The breakdown of CPU times for the $n_p = \{ 512-4096 \}$ series of jobs  is shown in Fig.~\ref{FIG_FMM_strongscaling}. 
The downward sweep event involves both a parallel operation (indicated by ``DownSweep'' in Fig.~\ref{FIG_FMM_strongscaling}) and a sequential operation at the root level of the tree (indicated by ``Root Tree DownSweep'' in Fig.~\ref{FIG_FMM_strongscaling}).
Thus, if the time required for the sequential operation is large compared to the time spent in evaluating contributions from the local parts of the tree, strong scalability will obviously suffer. The local calculations are all observed to strong scale well, however as the subdomains become smaller, the cost of the root tree will eventually dominate the overall execution time and reduce the parallel efficiency. In our experiments, the cost of the root tree evaluation grows by a factor of eight each time $r_l$ is increased by one. To offset the increasing cost of root level calculation, i.e. to observe better strong scalability, one can easily introduce work on each subdomain by increasing $\bar{M}$.
\begin{table}
	\caption{Strong scaling of \PetFMM on CADMOS BG/P. The times reported here represent the total time taken to perform the multipole summation (ParaFMMEvaluate). (S) denotes -mode SMP, (D) denotes -mode DUAL, (V) denotes -mode VN.
	$*$ indicates efficiency was computed w.r.t the 64 CPU execution time $(p_1=64)$. }
	\label{Table_FMM_strongscaling}
	\begin{tabular}{c c c c   c c}
	\hline 
	$\boldsymbol{n_p}$	&$\boldsymbol{r_l}$	&$\boldsymbol{L_v}$	&$\boldsymbol{\bar{M}}$		&\bf{CPU time (sec)}	 &\bf{Efficiency} \\
%	\hline 
%	\hline 
			&		&		&				&				&\\
	8		&2		&4		&96				&3.9740e+02 (S) 	& - \\
	16		&		&		&				&2.0950e+02 (S)	&95\%\\
	32		&		&		&				&1.1086e+02 (S)	&90\%\\
	64		&		&		&				&5.9088e+01 (D) 	&84\%\\
%	\hline 
			&		&		&				&				&\\
	32		&3		&5		&192			&9.2118e+02 (S)	&-,- \\
	64		&		&		&				&4.8627e+02 (D)	&95\%, - \\
	128		&		&		&				&2.5809e+02 (S)	&89\%,  94\%$^*$\\
	256		&		&		&				&1.4380e+02 (D)	&80\%,  85\%$^*$\\
	512		&		&		&				&8.6693e+01 (V)	&66\%,  70\%$^*$\\
%	\hline %% queue
			&		&		&				&				&\\
	512		&4		&6		&384			&7.8231e+02 (V) 	&-\\
	1024		&		&		&				&5.5052e+02 (D)	&71\%\\
	2048		&		&		&				&4.3421e+02 (D)	&45\% \\
	4096		&		&		&				&3.7705e+02 (V)	&26\% \\
	\hline 
	\end{tabular}
\end{table}

% ====================================================== %
\section{Discussion \label{SEC_Discussion}}

%	Errors
	In the experiments described in Sec.~\ref{SSEC_DiscretisationError}, the discretization error of the three methods was examined. In the norms measured, the convergence rates obtained using \SUMGone and \SUMGtwo were nearly identical. A measurable difference in the absolute error between the different quadrature rules was observed, with \SUMGone yielding errors approximately 2.3 times larger than \SUMGtwo. The rates measured between the two summation methods using Gauss quadrature 
and the rates obtained using \PetFMM, were extremely similar, provided the expansion order $p$ was high enough. In the cases where $p \le 4$, sub-optimal convergence $(E_2)$, or divergence was observed ($E_1,E_\infty$).
	
	Both \PetFMM and the summation methods incorporate the analytic solution of the potential (or gravity) within the discretization, thus these methods naturally satisfy the boundary condition, $\phi = 0$, as $\vtr x \rightarrow \infty$. Within the FE methods considered here, this boundary condition was approximated. 
The convergence behavior of the gravity field obtained using finite element methods is thus likely to be dependent on the choice of approximation made.	
In the absence of any boundary condition approximation and any approximations in defining the density structure, we anticipate the gravity error computed with $Q_1$ elements to behave like
\begin{equation}
	\| \vtr g - \vtr g^h \|_2  \le c_1 h,
	\label{eq:gravity_error}
\end{equation}
where $c_1$ is a constant independent of the grid resolution $h$.
 In the case of \FEMD, the boundary condition approximation is seen to limit how close the discrete solution will approximate the exact solution. Since the approximate boundary condition doesn't approach the true boundary condition in the limit of $h \rightarrow 0$, the convergence of rate of the potential and gravity field will ultimately deteriorate with increasing grid resolution. That is we have,
\begin{equation}
	\| \vtr g - \vtr g^h \|_2  \le c_1 h + c_2 \left( \frac{H}{L} \right).
\end{equation}
This type of relationship is evident in Fig.~\ref{FIG_FEM_error} where we observe a low correlation between the straight line with slope $0.57$ and the measured error. In practice this effect can be reduced if we ensure that the model domain is significantly larger than the domain defining the density anomaly, thereby making the coefficient $c_2$ smaller. However, adopting this approach introduces significantly higher computational requirements. 
	
	On the contrary, the alternative boundary condition approximation used in \FEMGT does not appear to place a bound on the minimum discretization error possible on a finite sized domain. This is apparent from Fig.~\ref{FIG_FEM_error} where a high correlation between the grid size and discretization error is observed. This suggests that the Robin boundary approximation converges like $\mathcal O(h)$ as the mesh is refined, since we observe the first order convergence predicted from Eq.~\eqref{eq:gravity_error} in the gravity field and this convergence rate appears to be independent of the domain aspect ratio $L/H$ (See Fig.~\ref{FIG_FEM_convergence}). Nevertheless, despite the improved convergence rate of \FEMGT, the rates observed are lower than those obtained using either the summation methods or \PetFMM.
	
%	Solve time
	To assess the speed of the three methods examined, we consider defining the cross over point where the summation methods cease to be less efficient than either \FEMGT and \PetFMM. The cross over point occurs when the number of evaluation points exceeds $t_{\text{\FEMGT,\PetFMM}} / t_\text{sum}$, where $t_\text{sum}$ is the time per evaluation point obtained from one of the summation algorithms. The number of evaluation points required to reach the cross over point for the sequential results are presented in Table.~\ref{Table_crossover}. We note that the times from Table~\ref{Table_SUM_timings} are repeated in the second and third column. The summation methods were not run at a grid resolution of $\bar{M} = \{192, 384 \}$, therefore the time required for the summation algorithms was estimated from the time required by the $\bar{M} = 96$ case and scaling this value by 8 and 64 respectively. 
%\hl{Need to think about these results a little more...}
\begin{table}
	\caption{Cross over point between the summation algorithms and the PDE based approaches. The rows marked with the $*$ indicate that the summation times were estimated from the summation simulation with $\bar{M}=96$. Columns 4-7 indicate the number of evaluation points below which the summation algorithms are faster than \FEMGT and \PetFMM.}
	\label{Table_crossover}
	\begin{tabular}{c c cc c cc cc}
	\hline 
	$\boldsymbol{\bar{M}}$		&\multicolumn{2}{c}{\bf{CPU time (sec) / station}}		&&\multicolumn{2}{c}{\bf{\FEMGT}}			&\multicolumn{2}{c}{\bf{\PetFMM}} \\	
			&\bf{\SUMGone}$\boldsymbol{(z)}$	&\bf{\SUMAN}				&&\bf{\SUMGone}$\boldsymbol{(z)}$	&\bf{\SUMAN}				&\bf{\SUMGone}$\boldsymbol{(z)}$	&\bf{\SUMAN} \\
%	\cline{1-3} 
%	\cline{5-8} 
			&			&					&&			&					&			& \\
	12		&3.71e-06	&5.50e-04			&&2.88e+04	&1.95e+02			&2.16e+04	&1.46e+02 \\
	24		&2.98e-05	&4.39e-03			&&3.39e+04	&2.30e+02			&3.99e+04	&2.71e+02 \\
	48		&2.37e-04	&3.76e-02			&&3.41e+04	&2.15e+02			&5.65e+04	&3.56e+02 \\
	96		&1.90e-03	&2.83e-01			&&3.39e+04	&2.28e+02			&6.68e+04	&4.49e+02 \\
	$192^*$	&1.52e-02	&2.26e+00			&&3.08e+04	&2.07e+02			&7.30e+04	&4.90e+02 \\
	$384^*$	&1.22e-01	&1.81e+01			&&3.08e+04	&2.06e+02			&7.67e+04	&5.15e+02 \\
	\hline 
	\end{tabular}
\end{table}
		
%	Parallelism
	All the three methods were observed to exhibit good strong scaling up to 1024 CPUs. By far the easiest method to obtain good parallel scalability was the summation methods. This is simply due to the lack of algorithmic complexity in the direct summation approach. Scalability here is only limited by the network of the computer cluster used. Our tests were performed on an IBM Blue Gene/P, which is known to have an excellent network with specialized hardware for performing global reductions. The techniques used by \FEMD,\FEMGT  and \PetFMM are more difficult to obtain high strong scaling efficiency. In the context of the multigrid preconditioner, this was due to the design choice that the mesh on the coarse grid had to be distributed and that we required at least one element per CPU. This particular restriction could be relaxed if a different coarse grid solver was employed. For example, we could use a large coarse grid, use less levels in the preconditioner and employ an exact coarse grid solve using an algebraic multigrid (AMG) preconditioner. The AMG algorithms are useful in this context as they do not require any geometric information to determine how the work will be distributed across the CPUs. With \PetFMM, speedup was measured up to 4096 CPUs, however the measure efficiency was only 26\%. Strong scaling with \PetFMM is hindered by the sequential calculations which have to be performed at the root level. This is a typical bottleneck in FMM algorithms,
however it could be eliminated by overlapping the root tree computation with the local direct summation work. This will be the object of future research.

%	Robustness and usability
Lastly we consider the overall usability of the different methods for computing gravity anomalies from the perspective of an end user. The quadrature based methods are by far the easiest method to use. It permits complete geometric freedom in defining the underlying grid which is used to discretize the density field. No connectivity is required between the cells and the vertices. The only requirement is that the cells used to partition the domain defined by the density anomaly do not overlap. Consequently, topography, curvature and locally refined regions are easily introduced. In the method described here, a constant density was used within each cell. This is not strictly necessary and spatial variations of density within a cell are possible, however the order of the quadrature rule used would likely have to be increased to maintain the accuracy of the method. 

	To use the geometric multigrid, a mesh hierarchy is required. Here we considered nested hierarchies of structured meshes. With such a topology, generating a mesh which has element faces which conform to all the jumps in density may be difficult to construct. This could be partially alleviated by using an unstructured mesh, but fully unstructured meshing in parallel is still a challenging task. Furthermore, an unstructured mesh hierarchy would also be required to be generated. The convergence, and hence the CPU time required by the geometric multigrid method is strongly dependent on the mesh geometry. For example, the implementation described here ceases to be robust if the mesh possesses a high aspect ratio, or the elements are highly deformed. In such circumstances, stronger smoothers are required if rapid convergence is to be maintained. Stronger smoothers may for example include block Jacobi with ILU factorization defined on the sub-blocks. Such choices mandate additional storage and careful selection and tuning of smoothers to remain optimal. 
%The choice of discretisation is also somewhat restricted if we wish to only use geometric multigrid. 
%#Fully unstructured geometric multigrid is possible, but defining a grid hierarchy in 3D and in parallel is still a challenging problem. 
%\hl{In our geometric multigrid implementation, we require for a given grid, that each processor's local subdomain must contain at least one element. This places certain restrictions on the size of the coarse grid which we can use for a given number of CPU's.}
To some extent, many of the aforementioned disadvantages related to geometric restriction introduced by using GMG can be overcome using algebraic multigrid (AMG). AMG preconditioners require the stiffness matrix to be assembled and furthermore, maintaing both scalable and optimal solution times in parallel is still a challenge with these approaches. The FE approaches does have the advantage that continuous density variations can be naturally introduce throughout the element.

The FMM does not possess any geometric restrictions in how the density structure may be defined. Whilst structured grids were used here, FMM can in principal be used with a random point distribution which define the location and value of density in space. In the case when a random distribution of points is used, one also needs to provide the volume of the domain which is associated to each point. This can be readily computed using a Voronoi diagram, or preferentially in parallel calculations using an approximate Voronoi diagram. Thus the method provides high geometric fidelity without having the burden of creating a mesh, conforming or otherwise. 
The time required to compute the gravity signal is a function of the number of points used to discretize the density, and not dependent on their spatial distribution. 
The convergence of FMM could be improved by introducing a basis with more smoothness than the current delta function discretization. In future work, we will introduce a Gaussian basis for rock masses so that the convergence rate can be adjusted by varying the width of the Gaussians. The initial interpolation problem for this new basis will be solved using the \PetRBF code~\citep{YokotaBarbaKnepley10}.

Despite being more than two times slower than \FEMGT, we believe that the geometric flexibility permitted in defining the density structure, combined with fact that the solve time is independent of the geometry of the discretization used for the density structure, make \PetFMM more useful in applied geophysics studies. \PetFMM was shown to be comparable in accuracy to the \SUMGtwo algorithm and should be used preferentially over this method if the number of evaluation points exceeds $77\times10^4$.

% ====================================================== %
\section{Conclusion}

Fast and robust forward models for computing a gravity signal from a density distribution is essential to perform high resolution inversions of the density subsurface.
Here we have discussed three different forward modes for computing gravity anomalies and compared them based on the convergence rates of the obtained gravity field, the execution time required to evaluate the gravity field and the parallel scalability of the algorithms. 
We considered classical summation techniques based on closed form expressions or quadrature schemes, and optimal and scalable approaches suitable for solving the Poisson equation. The PDE based approaches consisted of a finite element discretization utilizing a geometric multigrid preconditioner and an implementation of the fast multipole method.

The summation methods employing quadrature approximations are found to yield results of comparable accuracy to FMM. Only the finite element method which incorporated a far-field gravitational approximation in the form of a Robin boundary condition was deemed to be useful in practice. The error incurred by specifying a vanishing potential on the boundary of a finite domain resulted in large errors, and low convergence rates in the gravity field. All the forward models demonstrated good strong scaling up to 1024 CPUs. The fast multipole method presents itself as a viable alternative to classical summation methods due to the geometric freedom in defining the density structure and insensitivity of the overall CPU time to the underlying density structure. In comparison to the summation algorithm employing analytic expression for the gravity, FMM is faster provided more than 515 evaluation points are used. If the simplest quadrature based summation algorithm is used, FMM will provide a faster forward model if more than $77 \times 10^4$ evaluation points are used.
 
\subsubsection*{Acknowledgments}
The authors wish to thank Laetitia Le Pourhiet for computer time on {\it Octopus}.
All parallel computations were performed on the CADMOS BG/P, for which the authors thank Yuri Podladchikov.
%The authors thank the anonymous reviewers for their useful comments which greatly improved the clarity of the paper.
Author DAM was supported by the ETH Z\"urich Postdoctoral Fellowship Program. Partial support was provided by the
European Research Council under the European Community's Seventh Framework Program (FP7/2007-2013) / ERC Grant agreement
\#258830. MGK acknowledges partial support from NSF grant EAR-0949446.

% bibliography
%%\bibliographystyle{elsart-harv}
%%\bibliography{gravity_literature}

\appendix
\section{Error Evaluation} \label{SEC_ErrorEvaluation}
Here we discuss the method used to evaluate errors defined in Eqs.~\eqref{EQ_L1norm},\eqref{EQ_L2norm} and \eqref{EQ_Linfnorm}. The spatial variation of the discrete solution for the gravity field $g_z^h$ is defined by the representation natural to discretization. For the summation and FMM, this means $g_z^h$ is represented via piecewise constant over each cell. For the FE methods, $g_z^h$ is represented via a bilinear function $g_z^h = a_0 + a_1 x + a_2 y + a_3 xy$, since the potential $\phi$ was discretized via trilinear basis functions. The integrals in Eq.~\eqref{EQ_L1norm},\eqref{EQ_L2norm} were approximate via Gauss quadrature. 
The order of the quadrature used was determined empirically. The complexity of the analytic solution was such that low order rules were not appropriate to accurately estimate the norm. Over each cell in the discretization, we found that a 4-point quadrature rule, applied over $m \times m \times m$ subdivision (in each $x,y,z$ direction respectively) of each cell was sufficiently accurate. The value for $m$ was obtained by evaluating $\| g_z \|_1, \| g_z \|_2, E_\infty$ and examining how the error norm varied with $m$. The results from the experiment used to determine the value of $m$ for each $\bar{M}$ are presented in Table~\ref{Table_Appendix_quad_error}. The final value of $m$ shown for each $\bar{M}$ was used to calculate the norms in our experiments.
\begin{table}
	\caption{Results of the gravity quadrature test. Estimated values of the integral of the analytic gravity field obtained using a 4-point Gauss quadrature scheme, with different numbers of integration regions $m^3$, within each cell.}
	\label{Table_Appendix_quad_error}
	\begin{tabular}{c c c c c}
\hline 
$\boldsymbol{\bar{M}}$	&$\boldsymbol m$	&$\boldsymbol{\| g_z \|_1}$			&$\boldsymbol{\| g_z \|_2}$	&$\boldsymbol{E_\infty}$	\\
%\hline 
		&	&					&					&	\\
12		&3	&2.686359587701e+02	&3.461398542186e-02 &3.381867068310e-05	\\	
		&4	&2.686359587702e+02	&3.461399254307e-02 &3.403021478492e-05	\\	
		&5	&2.686359587700e+02	&3.461399453156e-02 &3.415713959000e-05	\\	
		&6	&2.686359587701e+02	&3.461399525775e-02 &3.424175549691e-05	\\	
		&7	&2.686359587703e+02	&3.461399557324e-02 &3.430219515320e-05 	\\	
		&8	&2.686359587699e+02	&3.461399572806e-02 &3.434752475888e-05 	\\	
%\hline 
		&	&					&					&	\\
24		&2	&2.686359587702e+02 &3.461399254307e-02 &3.403021478492e-05		\\
		&3	&2.686359587702e+02 &3.461399525775e-02 &3.424175549691e-05		\\	
		&4	&2.686359587698e+02 &3.461399572806e-02 &3.434752475888e-05 	\\	
%\hline 
		&	&					&					&	\\
48		&1	&2.686359587702e+02 &3.461399254307e-02 &3.403021478492e-05 	\\	
		&2	&2.686359587698e+02 &3.461399572806e-02 &3.434752475888e-05 	\\	
%\hline 
		&	&					&					&	\\
96		&1	&2.686359587698e+02 &3.461399572806e-02 &3.434752475888e-05		\\
\hline 
	\end{tabular}
\end{table}
The same quadrature rule used to evaluate $E_1,E_2$ was used to evaluate $E_\infty$.

\end{document}